\documentclass[twocolumn,showpacs,preprintnumbers,amsmath,amssymb,superscriptaddress]{revtex4-1}
\usepackage[dvipsnames]{xcolor}
\usepackage{threeparttable}
\usepackage{graphicx}
\usepackage{dcolumn}
\usepackage{bm}
\usepackage{tabularx}
\usepackage[utf8]{inputenc}
\usepackage[T1]{fontenc}
\usepackage[normalem]{ulem} 
\newcolumntype{M}{>{\centering\arraybackslash}m{1.85cm}}
\usepackage[export]{adjustbox}
\usepackage{float}
\usepackage{color}   
\usepackage{xurl}
\usepackage{hyperref}
\hypersetup{
    colorlinks=true, 
    linktoc=all,     
    linkcolor=blue,  
    breaklinks=true
}
\hypersetup{breaklinks=true}

\makeatletter
\newcommand{\colorcaption}[2][]{%
  \begingroup%
  \renewcommand{\@caption@fignum@sep}{ (Color online). }%
  \caption[#1]{#2}%
  \endgroup%
}
\makeatother

\newcommand{\orcid}[1]{\href{https://orcid.org/#1}{\hskip2pt\includegraphics[width=9pt]{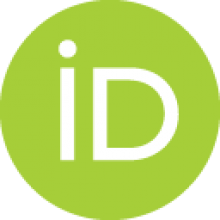}}}

\begin{document}

\title{A low-circuit-depth quantum computing approach to the nuclear shell model}

\author{Chandan Sarma}
\email{c.sarma@surrey.ac.uk}
\address{School of Mathematics and Physics, University of Surrey, Guildford, Surrey GU2 7XH, United Kingdom}	
\author{P. D. Stevenson}	
\email{p.stevenson@surrey.ac.uk}
\address{School of Mathematics and Physics, University of Surrey, Guildford, Surrey GU2 7XH, United Kingdom}

\date{\hfill \today}

\begin{abstract}
In this work, we introduce a new qubit mapping strategy for the Variational Quantum Eigensolver (VQE) applied to nuclear shell model calculations, where each Slater determinant (SD) is mapped to a qubit, rather than assigning qubits to individual single-particle states. While this approach may increase the total number of qubits required in some cases, it enables the construction of simpler quantum circuits that are more compatible with current noisy intermediate-scale quantum (NISQ) devices. We apply this method to seven nuclei:  Four lithium isotopes $^{6-9}$Li from the \textit{p}-shell, $^{18}$F from the \textit{sd}-shell, and two heavier nuclei ($^{210}$Po, and $^{210}$Pb). We run  circuits representing their ground states on a noisy simulator (IBM's \textit{FakeFez} backend) and quantum hardware ($ibm\_pittsburgh$).  For heavier nuclei, we demonstrate the feasibility of simulating $^{210}$Po and $^{210}$Pb as 22- and 29-qubit systems, respectively. Additionally, we employ Zero-Noise Extrapolation (ZNE) via two-qubit gate folding to mitigate errors in both simulated and hardware-executed results. Post-mitigation, the best results show less than 4 \% deviation from shell model predictions across all nuclei studied. This SD-based qubit mapping proves particularly effective for lighter nuclei and two-nucleon systems, offering a promising route for near-term quantum simulations in nuclear physics.


\end{abstract}


\maketitle
\section{Introduction}
\label{sect 1}

Since its introduction \cite{mayer_jensen}, the nuclear shell model has become a basic paradigm for representing and discussing structure properties of atomic nuclei \cite{shellmodelmdpi,suhonen}.  In its Configuration Interaction (CI) formalism, the shell model describes nuclear states through the mutual interaction of nucleons moving in a basis of single-particle orbitals.  The model's use of a relatively efficient basis means that nuclear wave functions can be represented by a manageable expansion in Slater Determinants of single particle states in the basis, at least for nuclei close to magic numbers.  Nevertheless, for some exotic states, or for nuclei far from magic numbers, the curse of dimensionality common to all many-body problems is encountered \cite{dean_computational_2008}.  Many innovative optimizations have been deployed in the implementation of the nuclear shell model over the years \cite{andreozzi_importance_2004,shimizu_thick-restart_2019,daonowacki,stumpf_importance-truncated_2016,launey_symmetry-guided_2016}, but ultimately the combinatorialy-increasing size of the Hilbert space with number of valence particles means that many nuclear systems will remain inaccessible to shell model calculations on classical computers.  

Quantum computation offers the promise of rendering large nuclear shell model problems tractable through the exponential scaling of multi-qubit Hilbert space with qubit number, and the possibility of efficiently representing highly-entangled states.  The shell model has become a model of choice for exploring nuclear structure problems with quantum algorithms \cite{QC_rev},  thanks in part to the simplicity in the $m$-scheme version of the shell model in mapping to qubit degrees of freedom, but also because it a basic paradigm which can describe simple test problems and large complex problems on the same footing \cite{kiss_quantum_2022,romero_solving_2022,stetcu,sarma_prediction_2023,perez-obiol_nuclear_2023,bhoy_shell-model_2024,hobday_variance_2025,li_deep_2024,carrascocodina2025comparisonvariationalquantumeigensolvers,nifeeya,costa_quantum_2025}.  As one realisation of a many-body quantum mechanics problem, the nuclear shell model also has much in common with other many-body problems, and one can hope to apply a common set of methods and to benefit from a cross-fertilization of ideas \cite{ayral_quantum_2023}.

Applying the nuclear shell model can mean different things, but a basic approach is to find the eigenstates of the shell model Hamiltonian, from which eigenstates observables can be calculated.  On current real quantum hardware, where noise and decoherence effects push practitioners towards low-depth circuits, the broad family of variational quantum algorithms (VQA) are widely used \cite{yuan_theory_2019,cerezo_variational_2022,fedorov_vqe_2022}.  As variational methods, they naturally target lowest energy states, but by various means such as preparing trial wave functions of particular symmetry, or pushing previously-found solutions to higher energy, a complete set of eigenstates can be found \cite{higgott_variational_2019,ssvqe,li_full_2024,hobday_variance_2025,choi_rodeo_2021, LCU}.

An art in the implementation of variational quantum algorithms is the development of suitable wave function anzatzes in circuit form.  Ideally the ansatz should be expressive enough to find the exact state desired, or a good enough approximation to it, while at the same time being simple enough and with few enough parameters that finding the lowest state of the cost function is achievable.  Considerable literature exists in developing suitable ansatzes either tailored to a problem at hand \cite{robin2025stabilizeracceleratedquantummanybodygroundstate,gibbs_exploiting_2025,bhoy_shell-model_2024,mihalikova_state_2025,costa2025quasiparticlepairingencodingatomic}, or remaining quite general while employing advanced optimization methods \cite{romero_solving_2022,robin_quantum_2023}.

In the present work, we recast the mapping between qubits and the shell model problem so that each qubit represents a Slater Determinant (SD) configuration rather than a single particle state, and show the conequences for variational quantum algorithms.  The driver is to produce simpler ansatzes with lower circuit depth, albeit at the expense of qubit number.  The expectation is that this method will be a viable alternative encoding of the nuclear shell model problem, appropriate for some current hardware where the number of available qubits exceeds typical shell model needs, but where error rates demand simple circuits. Another expected benefit of using this type of circuit is the future possibility of including three-nucleon forces within the same circuit design strategy.

\section{Formalism}
\label{sect 2}

\subsection{Hamiltonian}
The shell model Hamiltonian can be written in second quantization as
\begin{equation}
\label{eq1}
H = \sum_{i}\epsilon_i \hat{a}^\dagger_i  \hat{a}_i + \frac{1}{2} \sum_{i, j, k, l} V_{ijlk} \hat{a}^\dagger_i \hat{a}^\dagger_j \hat{a}_k \hat{a}_l.
\end{equation}

Here, the operators $\hat{a}^\dagger_i$ and $\hat{a}_i$ correspond to the creation and annihilation of a nucleon in the single-particle state $|i\rangle$. The parameters $\epsilon_i$ and $V_{ijlk}$ denote the single-particle energies and the two-body matrix elements (TBMEs), respectively. Each single-particle state is characterized by quantum numbers and can be written as $|i\rangle = |n, l, j, j_z, t_z\rangle$, where $n$ and $l$ represent the radial and orbital angular momentum quantum numbers. The quantum numbers $j$, $j_z$, and $t_z$ specify the total angular momentum, its projection along the $z$-axis, and the isospin projection, respectively. 

The $m$-scheme approach to the shell model \cite{deshalit_talmi,whitehead_numerical_1972} involves combining the single-particle basis states into all possible combinations consistent with a fixed $M=\sum j_z$ and fixed particle number for the nucleus of interest.  These combinations define the $m$-scheme basis in which the Hamiltonian (\ref{eq1}) is constructed, and then diagonalized to find the energy eigenvalues as well as eigenstates, from which other observables can be dervied.

This scheme is used extensively in contemporary shell model codes like KSHELL \cite{shimizu_thick-restart_2019}, NUSHELLX \cite{nushellx}, and BIGSTICK \cite{bigstick}. Although the number of single-particle states in a given model space may be modest, the dimension of the many-body basis grows rapidly with the number of nucleons.  Each shell model implementation deals with the increase of Hilbert space dimension with problem size in its own way, but ultimately the size of the many-body basis is a limiting factor in the $m$-scheme shell model.  

As mentioned in the introduction, a quantum computing approach has the potential to cope with the dimensionality problem through the natural exponential scaling of the multi-qubit Hilbert space.  However, as noted, circuit depths in the usual $m$-scheme to qubit mapping can be prohibitively large for current hardware, and it is the purpose of this work to explore an alternative encoding in which we re-express the Hamiltonian in a basis of Slater determinant (SD) configurations.

Considering $|m \rangle$ and $|n \rangle$ to be two possible SDs of a particular nucleus, then the many-particle matrix element between them can be represented as $H_{mn}$ = $\langle m| H | n\rangle$. The Hamiltonian in Eq. \ref{eq1} can be rewritten as

\begin{equation}
\label{eq2}
H = \sum_{m, n} H_{mn} \hat{A}^\dagger_m  \hat{A}_n
\end{equation}

Here $\hat{A}^\dagger_m$ and $\hat{A}_n$ are the creation and annihilation operators of the many-particle SDs $|m\rangle$ and $|n\rangle$ and $H_{mn}$ are the many-particle matrix elements. 

The shell model Hamiltonian defined in Eq. (\ref{eq1}) can be converted into the qubit Hamiltonian with the Jordan-Wigner (JW) \cite{JW} transformation using the mapping

\begin{eqnarray}
\hat{a}^\dagger_k = \frac{1}{2}  \left ( \prod_{j = 0}^{k-1} -Z_j \right)(X_k -iY_k), \label{eq3}\\
\hat{a}_k = \frac{1}{2}  \left ( \prod_{j = 0}^{k-1} -Z_j \right)(X_k + iY_k). \label{eq4} 
\end{eqnarray}

The Hamiltonian in Eq. (\ref{eq2}) can also be transformed by the JW transformation, but in this case the $Z$ gates can be omitted, as the parities among the single particle states are already accounted for in the SD configurations.  Hence, a qubit Hamiltonian for Eq. (\ref{eq2}) can be directly constructed as 

\begin{equation}
\label{eq5}
H_{qubit} = \sum_{m} H_{mm} \frac{(I_m -Z_m)}{2} + \sum_{m < n} H_{mn} \frac{(X_m X_n + Y_m Y_n)}{2}
\end{equation}

In this work, we use this SD basis to explore the qubitization and subsequent solution of shell model Hamiltonians in the form of Eq. (\ref{eq2}), comparing to the standard single-particle mapping of previous studies via the usual form (\ref{eq1}).  We consider nuclei from three different mass regions, and correspondingly, we need to consider three different shell model interactions for them. For the low mass $p$-shell nuclei, we consider the Ckpot interaction \cite{ckpot}, and for the $sd$-shell, we use the well-known USDB interaction \cite{usdb}. For the heaviest nuclei under consideration, $^{210}$Po and $^{210}$Pb, we consider the KHPE interaction \cite{khpe}.

\subsection{Initial state preparation and variational ansatz}

\begin{figure}
    \centering
    \includegraphics[scale = 0.75]{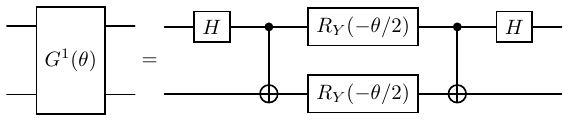}
    \caption{Single excitation Givens rotation in terms of basic quantum gates. Adapted from \cite{single_ex}.}
    \label{Single_ex}
\end{figure}

\begin{figure*}
    \includegraphics[scale = 0.6]{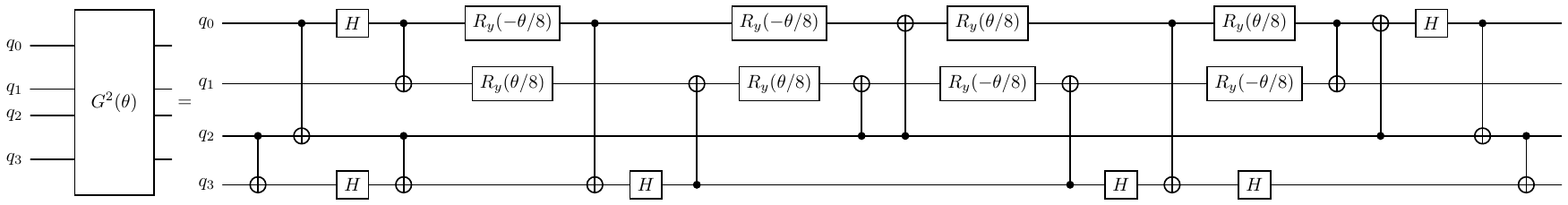}
    \caption{Double excitation Givens rotation in terms of basic quantum gates. Adapted from \cite{double_ex}.}
    \label{Double_ex}
\end{figure*}

For the SD basis, we have that each qubit represents a specific many-particle configuration. These configurations are linked by single excitation Givens roations thanks to the one-body expression of the Hamiltonian in (\ref{eq2}).  In the single-particle basis, the two-body form of the Hamiltonian in Eq. (\ref{eq1}) means that double excitations are needed and we use a general double Givens rotation to achieve this \cite{bhoy_shell-model_2024}.  The representation we use of the single and double Givens rotations in terms of more elementary quantum circuit gates is shown in \autoref{Single_ex} (single) and \autoref{Double_ex} (double).

To explain the initial state preparation for each mapping, we consider the case of the $^6$Li ground state.  This state is known to have spin $J$ = 1 and so the lowest energy $M=1$ state is appropriate to represent it.  

\begin{figure}
    \centering
    \includegraphics[width=0.47\textwidth, trim=1cm 1cm 1cm 3cm, clip]{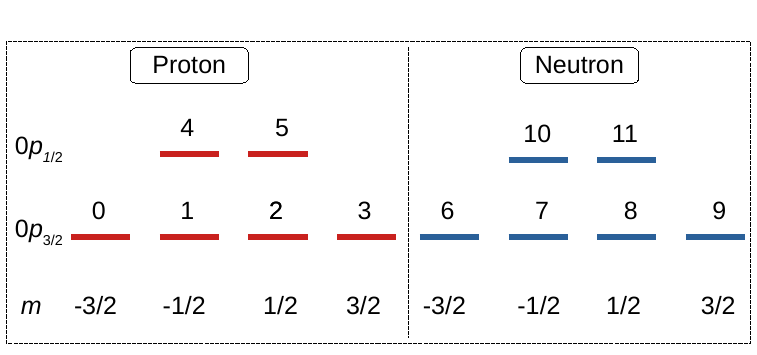}
    \caption{Single particle states of the $p$-shell.  $l, j, m$ quantum numbers given in usual nuclear notation.  The indexing from 0 to 11  are used for qubit mapping.}
    \label{pshell}
\end{figure}

\begin{figure*}
    \includegraphics[scale = 0.75]{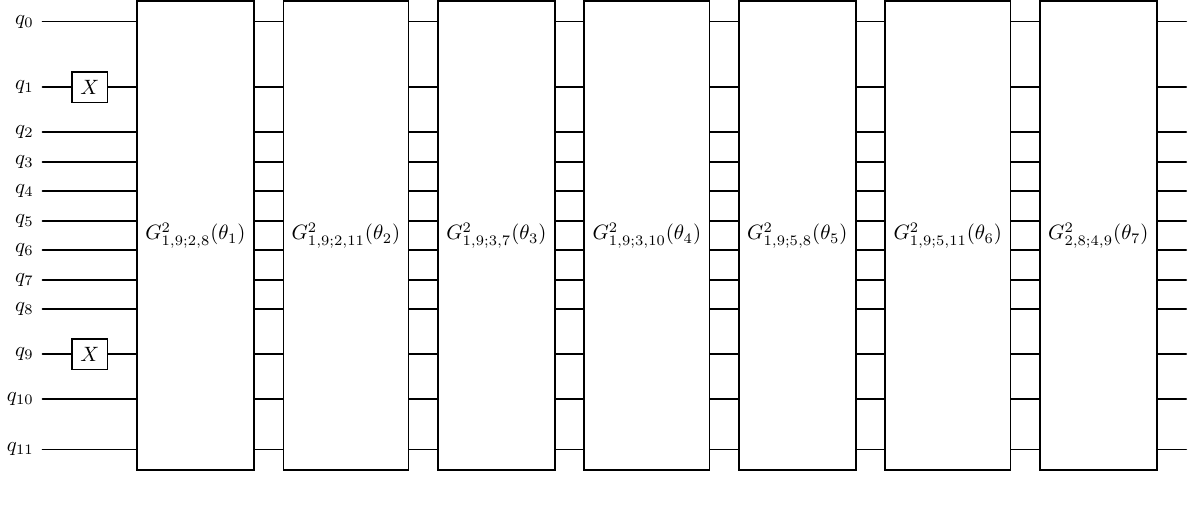}
    \caption{Double excitation ansatz for $^6$Li (1$^+$).}
    \label{6Li_UCCD}
\end{figure*}

In the single particle basis mapping, one takes each single particle in the basis shown in \autoref{pshell} and maps it to a qubit using the indexing of \autoref{pshell}, then initialises a circuit such that one possible $M=1$ configuration is activated through $X$ gates. From there, a series of double excitation Givens' rotations shares the amplitude of the wave function to all other possible $M=1$ combinations.  In \autoref{6Li_UCCD} a suitable circuit is shown for the prepration of an $M=1$ state with the single-particle mapping.  Here, the qubit indices correspond to the single particle level labels in \autoref{pshell} and the subscripts on the double excitation gates indicate the qubits on which the gates act: E.g. $G^2_{1,9;2,8}(\theta_1)$ rotates the qubit pair $1,9$ to the pair $2,8$ with a rotation angle $\theta_1$.  It is possible to ensure that some or all of the double excitations work on four neighboring qubits by relabeling the single-particle to qubit mapping \cite{bhoy_shell-model_2024}.  The effort of doing this is worthwhile for implementation on hardware with nearest-neighbor connectivity, but since we are using the single-particle mapping only a reference calculation in simulation, the connectivity issue is not relevant.

\begin{figure*}
    \centering
    \includegraphics[scale = 0.58]{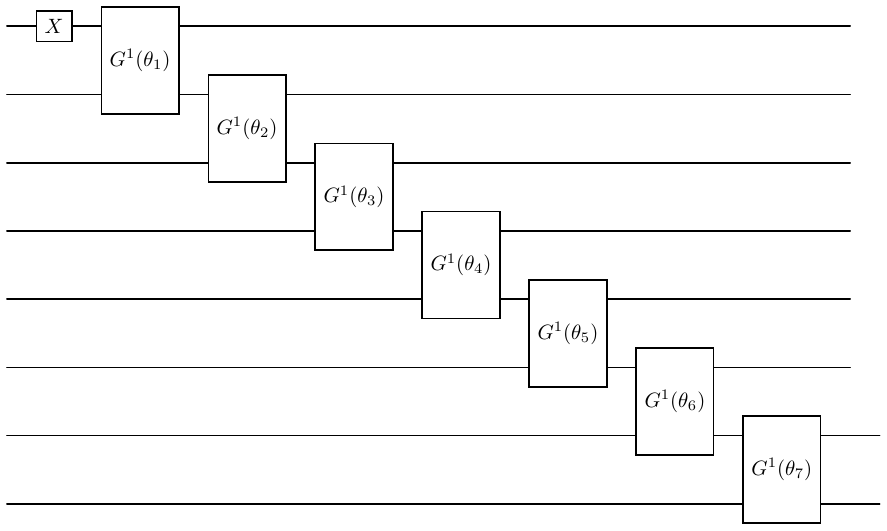}
    \caption{Single excitation ansatz for $^6$Li ($1^+$).}
    \label{6Li_single_ex}
\end{figure*}

In the SD basis, we first identify the configurations which have the appropriate $M$ value, and then identify each configuration with a qubit.  The eight $M=1$ configurations for a $J=1$ state in $^6$Li are enumerated in \autoref{Li_SD}.  Note that these eight configurations correspond to the initialsed states and the 7 subsequent double excitations in the single-particle circuit in \autoref{6Li_UCCD}.  The SD basis circuit needs to begin by initialising a single configuration with an $X$ gate, then using single excitations to spread amplitude across all other allowed configurations.  Such a circuit is shown in \autoref{6Li_single_ex}. The indexing of the qubit and mapping to the configurations is as given in \autoref{Li_SD}.

The determination of the angles in the Givens' rotations is made through a standard variational quantum eigensolver (VQE).  In the present work, we perform the VQE iterations in simulation to determine the angles, then evalute the resulting circuit with fixed angles on real quantum hardware.

\begin{table}
\caption{Qubit assignments based on the possible SDs for the ground state of lithium isotopes. The i, j, k for an SD |i, j, k$\rangle$ represent the single particle state indexed in \autoref{pshell}.}

    \centering
    \begin{tabular}{|c|c|c|c|c|}
    \hline
        Qubits &  $^6$Li (1$^+$) & $^7$Li (3/2$^-$) & $^8$Li (2$^+$) & $^9$Li (3/2$^-$)\\
    \hline    
         0 &  |1, 9$\rangle$ & |1, 8, 9$\rangle$ & |1, 8, 9, 11$\rangle$ & |1, 7, 8, 9, 11$\rangle$\\
         1 & |2, 8$\rangle$ & |1, 9, 11$\rangle$ & |2, 7, 8, 9$\rangle$ & |1, 8, 9, 10, 11$\rangle$ \\
         2 & |2, 11$\rangle$ & |2, 7, 9$\rangle$ & |2, 7, 9, 11$\rangle$ & |2, 6, 8, 9, 11$\rangle$ \\
         3 &  |3, 7$\rangle$& |2, 8, 11$\rangle$ & |2, 8, 9, 10$\rangle$ & |2, 7, 8, 9, 10$\rangle$ \\
         4 & |3, 10$\rangle$ & |2, 9, 10$\rangle$ & |2, 9, 10, 11$\rangle$ & |2, 7, 9, 10, 11$\rangle$  \\
         5 & |4, 9$\rangle$ & |3, 6, 9$\rangle$ & |3, 6, 8, 9$\rangle$ & |3, 6, 7, 8, 9$\rangle$ \\
         6 & |5, 8$\rangle$ & |3, 7, 8$\rangle$ & |3, 6, 9, 11$\rangle$ & |3, 6, 7, 9, 11$\rangle$ \\
         7 & |5, 11$\rangle$ & |3, 7, 11$\rangle$ & |3, 7, 8, 11$\rangle$ & |3, 6, 8, 9, 10$\rangle$ \\
         8 & -- & |3, 8, 10$\rangle$ & |3, 7, 9, 10$\rangle$ & |3, 6, 9, 10, 11$\rangle$ \\
         9 & -- & |3, 10, 11$\rangle$ & |3, 8, 10, 11$\rangle$ & |3, 7, 8, 10, 11$\rangle$ \\
         10 & -- & |4, 8, 9$\rangle$ & |4, 8, 9, 11$\rangle$ & |4, 7, 8, 9, 11$\rangle$ \\
         11 & -- & |4, 9, 11$\rangle$ & |5, 7, 8, 9$\rangle$ & |4, 8, 9, 10, 11$\rangle$ \\
         12 & -- & |5, 7, 9$\rangle$ & |5, 7, 9, 11$\rangle$ & |5, 6, 8, 9, 11$\rangle$ \\
         13 & -- & |5, 8, 11$\rangle$ & |5, 8, 9, 10$\rangle$ & |5, 7, 8, 9, 10$\rangle$ \\
         14 & -- & |5, 9, 10$\rangle$ & |5, 9, 10, 11$\rangle$ & |5, 7, 9, 10, 11$\rangle$ \\
     \hline     
    \end{tabular}
    \label{Li_SD}
\end{table}

\begin{figure*}
    \centering
    \includegraphics[scale = 0.56]{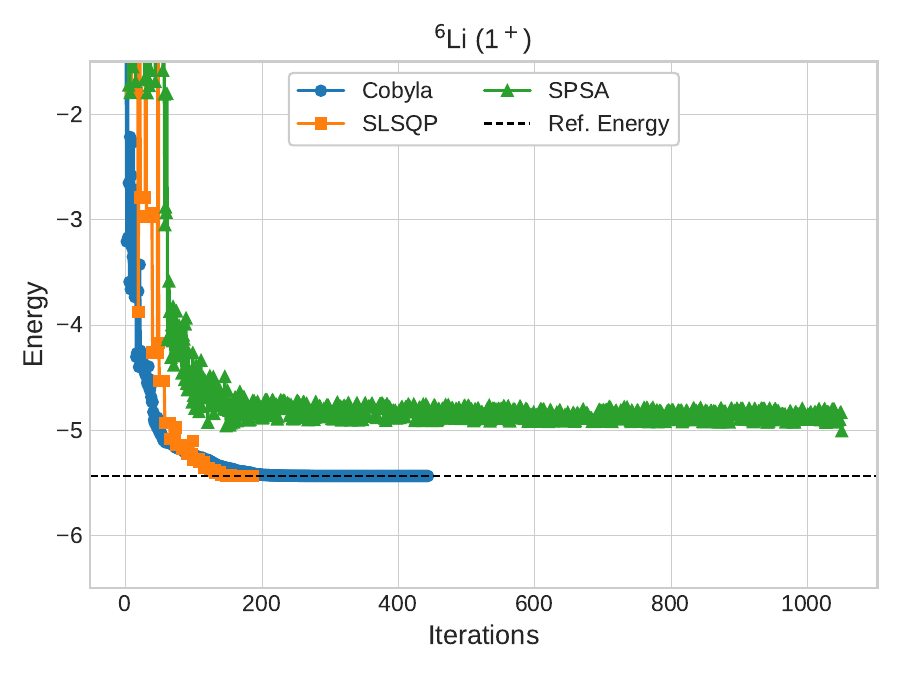}
    \includegraphics[scale = 0.56]{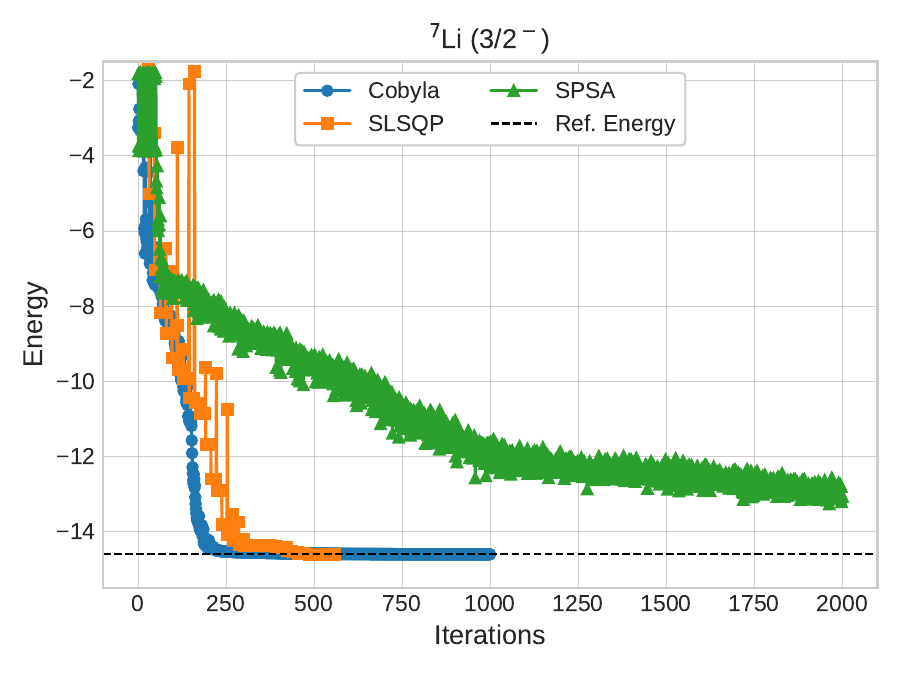}
    \caption{Convergence of binding energies of (a) $^6$Li (1$^+$) and (b) $^7$Li (3/2$^-$) with the number of iterations for three optimizers, namely, COBYLA, SLSQP, and SPSA.}
    \label{6Li_vqe_run}
\end{figure*}

\section{Results and Discussions}
\label{sect 3}

\subsection{Quantum simulation of Li isotopes}

We start by considering the  $^{6-9}$Li ground states using the SD basis.  In the previous section, the example of $^6$Li was considered, with the enumeration of the available configurations and the corresponding circuit of single excitations.  For the cases of isotopes with mass numbers 7,8,9, the number of valence particles in the $p-$shell is 3,4,5 respectively, and the allowed configurations of the 3,4,5-particle states need to be enumerated before a circuit ansatz for the nuclear wave function can be made.  

\autoref{Li_SD} shows the enumeration of the allowed configurations consistent with the $J=M$ value of the ground state for each of the four lithium isotopes under consideration. It is seen that the number of configurations, and hence qubits, needed for these nuclei is up to 15.  This is in contrast to the single-particle basis in which the 12 states shown in \autoref{pshell} gives the upper bound of qubits needed.


\begin{table*}
\caption{The ground state spins and parities of four Li-isotopes and their ground state reference (``Ref.'') energies from shell model calculations are shown along with the required resource counts to simulate them using the single excitation ansatz shown in \autoref{6Li_single_ex}. }

    \centering
    \begin{tabular}{|c|c|c|c|c|c|c|c|c|}
    \hline
        Nucleus ($J^\pi$)& Ansatz & Qubits & Parameters & Pauli terms & 1$Q$ gates & 2$Q$ gates & Depth & Ref. energy (in MeV)\\
    \hline    
    $^6$Li (1$^+$) & Single Ex. & 8 & 7 & 65 & 29 & 14 & 36 & -5.437 \\
     & Single Ex. (Transpiled) & 8 & 7 & 65 & 197 & 14 & 134 &  \\
     & Single Ex. (Optimized) & 8 & 7 & 65 & 113 & 14 & 67 &  \\
     \hline
    $^7$Li (3/2$^-$) & Single Ex.&15 & 14 & 180 & 57 & 28 & 71 & -14.607 \\
     & Single Ex. (Transpiled) &15 & 14 & 180 & 393 & 28 & 267 &  \\
     & Single Ex. (Optimized) &15 & 14 & 180 & 202 & 28 & 118 &  \\
     \hline
    $^8$Li (2$^+$) & Single Ex.& 15 & 14 & 182 & 57 & 28 & 71 & -14.926 \\
    & Single Ex. (Transpiled) & 15 & 14 & 182 & 393 & 28 & 267 &  \\
    & Single Ex. (Optimized) & 15 & 14 & 182 & 202 & 28 & 118 &  \\
    \hline
    $^9$Li (3/2$^-$) & Single Ex. &15 & 14 & 182 & 57 & 28 & 71 & -18.974 \\
     & Single Ex. (Transpiled) &15 & 14 & 182 & 393 & 28 & 267 &  \\
     & Single Ex. (Optimized) &15 & 14 & 182 & 202 & 28 & 118 &  \\
    \hline
              
    \end{tabular}
    \label{Li_sum}
\end{table*}

\begin{figure*}
    \centering
    \includegraphics[width=0.74\linewidth]{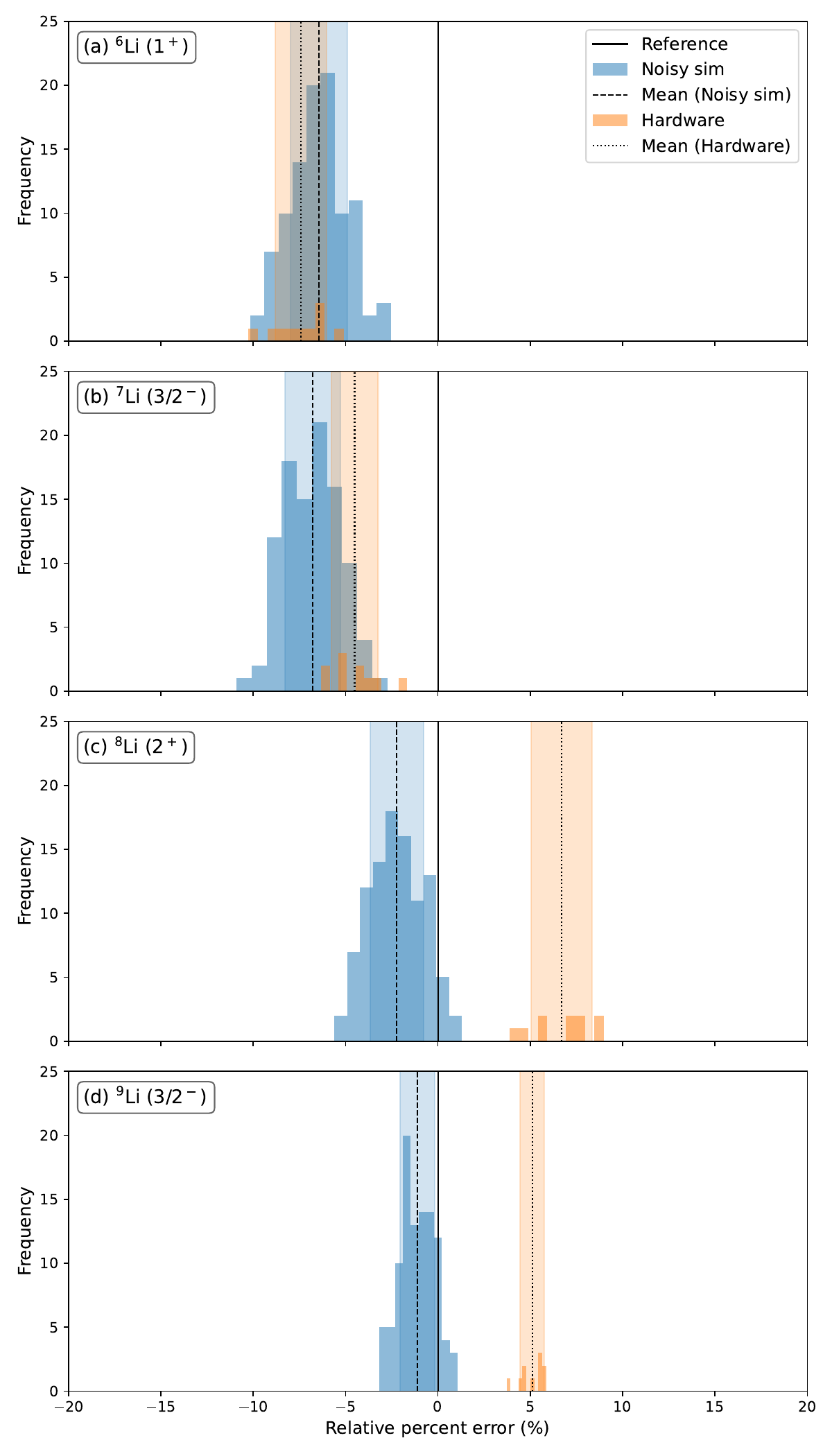}
    \caption{Histograms of single excitation circuits for $^{6-9}$Li.  In the ``Noisy sim'' case, 100 independent executions are made on the IBM \textit{FakeFez} backend using the optimal parameters from the noiseless simulation.  The ``Hardware'' results come from 10 independent executions on the IBM\_pittsburg device, using the noiseless optimal parameters.}
    \label{Li_histograms}
\end{figure*}

The payoff for the increased number of qubits is that every circuit has the simple form of a ladder of single excitations analagous to \autoref{6Li_single_ex} for all isotopes.  A resource count for the calculations is detailed in \autoref{Li_sum}.  For each isotope, resource counts are presented as ``Single Ex.'' for the SD basis single exictation ansatz as described above with the breakdown of the single excitation Givens rotation in terms of Hadamard, CNOT, and rotations gates as given in \autoref{Single_ex}.  Since our goal is to implement on real quantum hardware, we transpile our circuits with the IBM \textit{FakeFez} backend to analyze the gate count as used with a machine's native gate set (IBM\_Fez, whose gate set is $CZ$, $I$, $R_x$, $R_z$, $R_{zz}$, $\sqrt{X}$, $X$).  The gate count is show in the lines ``Single Ex. (transpiled)'' in \autoref{Li_sum}.  Reduction in native gate count given by level three optimization within the IBM compiler is then shown in the line ``Single Ex. (optimized)''.  As well as gate counts, the number of qubits needed, adjustable parameters in the variational ansatz, and number of terms in the representation of the Hamiltonian in terms of Pauli strings are also shown. 

 From the table, it can be seen that the resource counts for $^{7-9}$Li are almost the same since all three isotopes are defined as 15-qubit systems whose ground state ansatzes are constructed with 14-single excitations. 

For the determination of the optimized ground states, we consider the original ansatzes before transpilation and perform VQE runs using three different optimizers: Cobyla \cite{cobyla}, SLSQP \cite{slsqp} and SPSA \cite{spsa}. {\color{black} The optimizer codes are taken from the \texttt{Qiskit} library \cite{Qiskit} and the convergence of the binding energies with number of iterations are shown in \autoref{6Li_vqe_run} for $^6$Li and $^7$Li. From the figure, it can be seen that the Cobyla optimizer shows faster convergence, reaching the exact ground state binding energies in around 250 iterations. A similar trend is also observed in the case of $^8$Li (2$^+$) and $^9$Li (3/2$^-$) and based on this observation, we decided to consider the optimum parameters from Cobyla optimizers for these four Li-isotopes for evaluation of the circuit on real hardware.}

Having obtained optimized gate parameters for each Li isotope from a Cobyla minimization, we {\color{black}transpile those circuits and} run each 100 times independently using IBM's \textit{FakeFez} backend.  This is a  simulated backend based on the noise model from the 156-qubit $ibm\_fez$ quantum computer. The results are shown in \autoref{Li_histograms} {\color{black} and the relative percentage error is calculated as ($BE_\text{noisy sim} - BE_\text{exact})/BE_\text{exact}\times100$), where $BE_\text{noisy sim}$ and $BE_\text{exact}$ are the binding energies from noisy simulation and shell model, respectively}. From the figure, it can be seen that the noisy simulation results for $^6$Li (1$^+$), which has a eight qubit circuit are comparable to  $^7$Li (3/2$^-$), which has a fifteen qubit circuit both showing around 7 \% underbinding compared to the reference energy from the shell model calculation. 
On the other hand, for the neutron-rich Li-isotopes ($^8$Li and $^9$Li), the noisy simulated results are only 2.22 and 1.46 \% less bound than the reference binding energies.  We conjecture that it is the large component of the initialised qubit 0-mapped SD configuration making up the exact solution in the case of $^8$Li and $^9$Li that are the cause of the lower errors.  In these cases a greater part of the overall qubit wave function is created by the initial one-qubit $X$ gate, and less by the more error-prone two-qubit gates in the subsequent series of Givens rotations.


Finally, we run these circuits on the $ibm\_pittsburgh$ quantum computer for 10 independent runs, considering the same sets of optimal parameters used for the noisy simulation. The $ibm\_pittsburgh$ machine is also a 156 qubit quantum computer just like $ibm\_fez$ with the same set of hardware native gates. The results are shown in Fig. \ref{Li_histograms}, and from the figure, it can be seen that the hardware results are close to the noisy simulated results for $^6$Li and $^7$Li and show a similar underbinding. In contrast to the noisy simulation, the hardware results for the other two Li isotopes ($^8$Li and $^9$Li) show overbinding rather than underbinding as shown in Fig. \ref{Li_histograms}. Overall the mean binding energies from the hardware are of order 5 \% away from the shell model results and it shows the need to implement error mitigation techniques. 

\begin{figure*}
    \centering \includegraphics[width=0.77\linewidth]{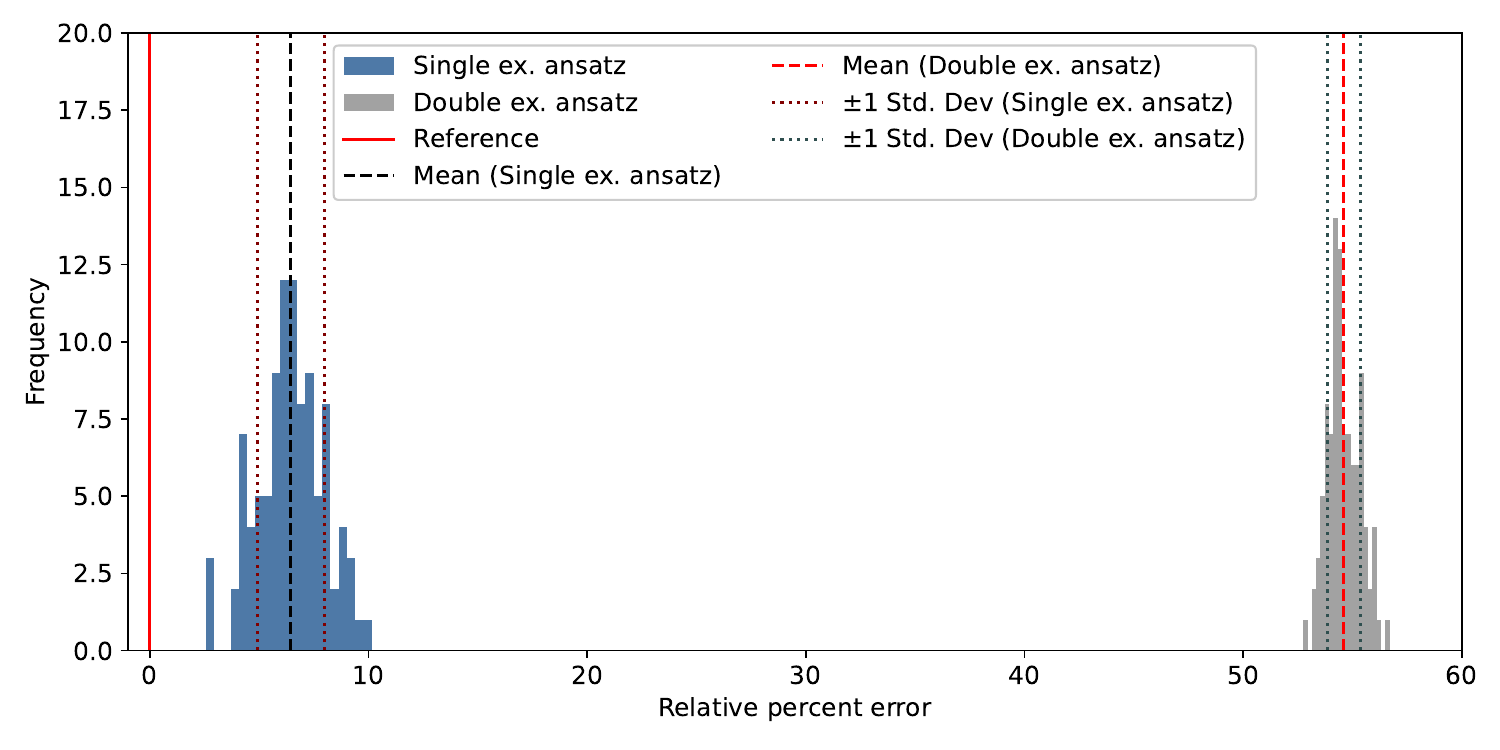}
    \caption{Comparison of the performance of single and double excitation circuits in reproducing the g.s binding energy of $^6$Li in noisy simulation.}
    \label{6Li_com}
\end{figure*}

\subsection{Comparison of single particle and Slater Determinant mappings for $^6$Li and $^{18}$F}

In this section, we compare the SD mapping with the single excitation ansatz, and the single-particle mapping with the  double excitation ansatz, considering two systems each with one proton and one neutron in the valence space: $^6$Li and $^{18}$F. As discussed in the previous section, the ground state of $^6$Li is defined as an 8-qubit system when considering the Slater determinant mapping. On the other hand, 12 qubits are used for the single-particle mapping. Similarly, the SD and single-particle mappings for the ground state of $^{18}$F can be implemented using 25 and 24 qubits, respectively. A comparison between these two approaches is given in \autoref{6Li_18F_sum}. From the table, it can be seen that the SD picture with its single excitation ansatz requires significantly lower resources in the ansatz as well as a lower number of Pauli terms to measure. It resulted in fewer errors in the noisy simulation or hardware results and required less execution time. A comparison between the performance of the two approaches is shown in \autoref{6Li_com} for 100 independent executions of the optimized circuits on the \textit{FakeFez} backend. Additionally it contains the results of 10 independent excutations on $ibm\_pittsburgh$ quantum computer. From the figure, it can be noted that while the single excitation results are less than 10 \% off from the exact results, the double excitation results show around 55 and 65 \%  underbinding for noisy simulation and hardware run, respectively. Though such a comparison is not shown for the case of $^{18}$F due to relatively long execution time, a similar conclusion can be expected as the $^6$Li case.  

\begin{table*}
\caption{The ground state spins and parities of four lithium isotopes and their reference (``Ref.'') energies from shell model calculations along with the required resource counts to simulate them using the single excitation ansatz shown in \autoref{6Li_single_ex}. }

    \centering
    \begin{tabular}{|c|c|c|c|c|c|c|c|c|}
    \hline
        Nucleus ($J^\pi$)& Ansatz & Qubits & Parameters & Pauli terms & 1$Q$ gates & 2$Q$ gates & Depth & Ref. energy (in MeV)\\
    \hline \\[-8pt]   
$^6$Li (1$^+$) & Single Ex. & 8 & 7 & 65 & 29 & 14 & 36 & -5.437 \\
& Single Ex. (Transpiled) & 8 & 7 & 65 & 197 & 14 & 134 &  \\
& Single Ex. (Optimized) & 8 & 7 & 65 & 113 & 14 & 67 &  \\
& Double Ex. & 12 & 11 & 207 & 100 & 98 & 132 & \\
& Double Ex. (Transpiled) & 12 & 11 & 207 & 1776 & 488 & 1027 & \\
& Double Ex. (Optimized) & 12 & 11 & 207 & 753 & 227 & 530 & \\
    \hline \\[-8pt]
$^{18}$F (1$^+$) & Single Ex. & 25 & 24 & 618 & 97 & 48 & 121 & -13.413 \\
& Single Ex. (Transpiled) & 25 & 24 & 618 & 919 & 171 & 589 &  \\
& Single Ex. (Optimized) & 25 & 24 & 618 & 371 & 48 & 213 &  \\
& Double Ex. & 24 & 23 & 2112 & 308 & 336 & 463 & \\  
& Double Ex. (Transpiled) & 24 & 23 & 2112 & 6848 & 2055 & 3758 & \\
& Double Ex. (Optimized) & 24 & 23 & 2112 & 3096 & 1003 & 2055 & \\
\hline
    \end{tabular}
    \label{6Li_18F_sum}
\end{table*}

\subsection{Quantum simulation of $^{210}$Po and $^{210}$Pb}

In this section, we discuss two heavier mass nuclei, $^{210}$Po and $^{210}$Pb, beyond $^{208}$Pb using the SD mapping and the single excitation ansatz with the SD mapping. We use the KHPE shell model interaction for this nucleus \cite{khpe}, having 44 proton single particle states and 58 neutron single particle states. The number of Slater determinants needed (i.e. the $m$-scheme dimensions) for the g. s. of $^{210}$Po and $^{210}$Pb are 62 and 99, respectively.  Since we are using simulators to obtain the optimized angles,  we are unable to consider such a large number of qubits.  However, for ground states, it is expected from the nuclear physics that we need only consider  those SDs which are the combinations of two time-reversed single particle states having the same $|j_z|$ with opposite signs. By doing so, the ground state of $^{210}$Po can be represented as a 22-qubit ansatz. Similarly, the ground state of $^{210}$Pb can be simulated using a 29-qubit ansatz.  A similar consideration was earlier made for some low to mid-mass two-nucleon systems in \cite{kimura}. The resource counts required to simulate these two nuclei are shown in \autoref{210Po_sum}. The average binding energy of $^{210}$Po obtained from noisy simulation is -9.438 MeV, which is around 8 \% more than the reference binding energy from shell model. However, due to a long execution time, noisy simulation is not done for $^{210}$Pb, instead we directly run the optimized circuits on $ibm\_pittsburgh$ device. The average ground state binding energies of $^{210}$Po and $^{210}$Pb as obtained from the hardware are -11.244 and -16.830 MeV, which are 28 and 85 \% more bound than the exact shell model results. As the $^{210}$Pb is the largest system considered in this work, {\color{black} with correspondingly deep circuits,} it is more prone to hardware error. 

\begin{table*}
\caption{The ground state spins and parities of $^{210}$Po and $^{210}$Pb and their energies from reference (``Ref.'') shell model calculations, along with the required resource counts to simulate them using single excitation ansatzes. }

    \centering
    \begin{tabular}{|c|c|c|c|c|c|c|c|c|}
    \hline
        Nucleus ($J^\pi$)& Ansatz & Qubits & Parameters & Pauli terms & 1$Q$ gates & 2$Q$ gates & Depth & Ref. energy (in MeV)\\
    \hline \\[-8pt]   
$^{210}$Po (0$^+$) & Single Ex. & 22 & 21 & 485 & 85 & 42 & 86 & -8.762\\
& Single Ex. (Transpiled) & 22 & 21 & 485 & 823 & 159 & 538 & \\
& Single Ex. (Optimized) & 22 & 21 & 485 & 325 & 42 & 191 & \\
    \hline \\[-8pt]
$^{210}$Pb (0$^+$) & Single Ex. & 29 & 28 & 842  & 113 & 56 & 114 & -9.091 \\
& Single Ex. (Transpiled) & 29 & 28 & 842 & 1061 & 194 & 683 &  \\
& Single Ex. (Optimized) & 29 & 28 & 842 & 423 & 56 & 246 &  \\
\hline
    \end{tabular}
    \label{210Po_sum}
\end{table*}

\subsection{Error mitigation}

In order to improve the results obtained from the noisy simulation and quantum hardware, we implement the zero noise extrapoltation (ZNE) technique as our choice of error mitigation technique \cite{ZNE, dumitrescu_cloud_2018}. For that we consider the two-qubit gate error mitigation using two-qubit gate folding. The two-qubit gate involved with the \textit{FakeFez} backend and with the $ibm\_pittsburgh$ quantum computer is the $CZ$ gate.  A single two-qubit gate fold involves adding a pair of $CZ$ gates acting as a identity. Then, standard extrapolation is carried out using noise scale factors of the form 2$\lambda$ + 1, where $\lambda$ represents the number of two-qubit folds and increases proportionally with the number of additional $CZ$ gates. The numerical values of circuit executions at different noise factors and extrapolated results are shown in \autoref{tab_zne}). For each nucleus, the first row shows the results of noisy simulation while the second row shows the hardware results. 

\begin{figure*}
    \centering 
    \includegraphics[scale = 0.65]{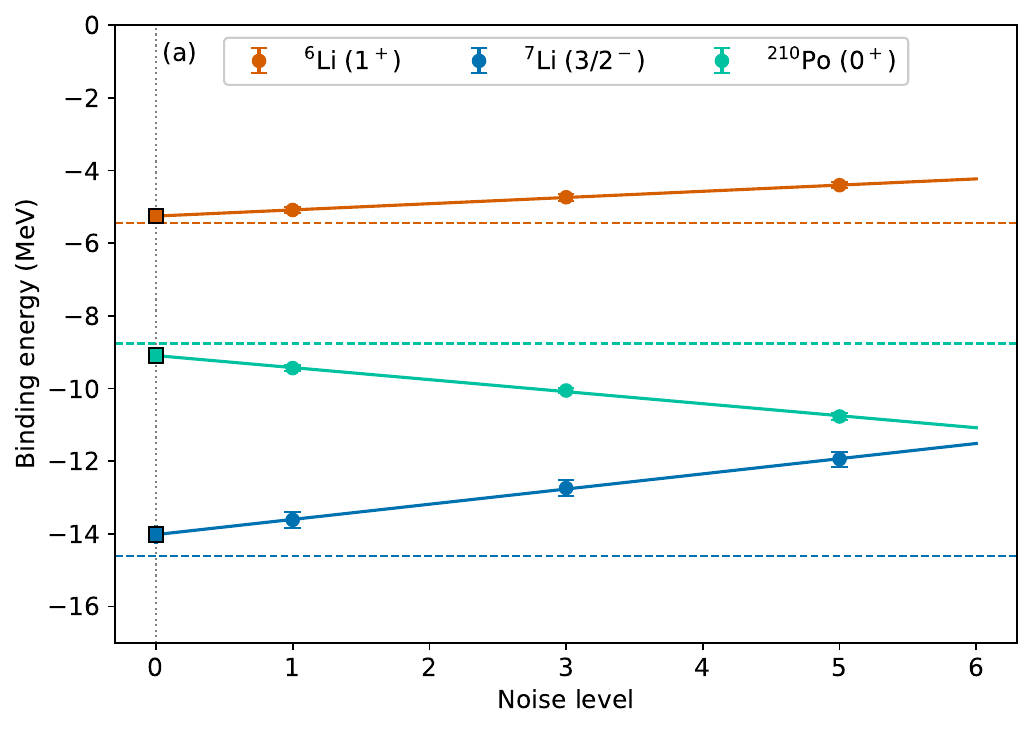}
    \caption{Zero noise extrapolation performed on the noisy simulated results for $^{6, 7}$Li and $^{210}$Po using linear extrapolation.}
    \label{zne_fig}
\end{figure*}

The ZNE error mitigation technique is first performed for $^{6, 7}$Li and $^{210}$Pb nuclei under noisy simulation. For these three nuclei, we evalaute the expectation values of the qubit Hamiltonians using the optimized circuits (noise scale 1), optimized circuits with single $CZ$ gate folding (noise scale 3) and optimized circuits with double $CZ$ gate folding (noise scale 5). For the Li isotopes each optimized circuit is executed 100 times independently, while for $^{210}$Po 10 independent run were executed and the final results are shown in the third, fourth and fifth columns of \autoref{tab_zne}. Now, using these three sets of results, we extrapolated these results to zero-noise limit using a linear extrapolation, a second-order polynomial extrapolation and an exponential extrapolation out of which the linear extrapolation is shown in \autoref{zne_fig}. The final ZNE results corresponding to linear, second degree polynomial and exponential extrapolation are shown in the sixth, seventh and eighth columns of \autoref{tab_zne}, respectively. Finally, in the last column the percent error corresponding to the best error mitigated values (bold-faced values in \autoref{tab_zne}) are shown. 

\begin{figure*}
    \centering 
    \includegraphics[scale = 0.65]{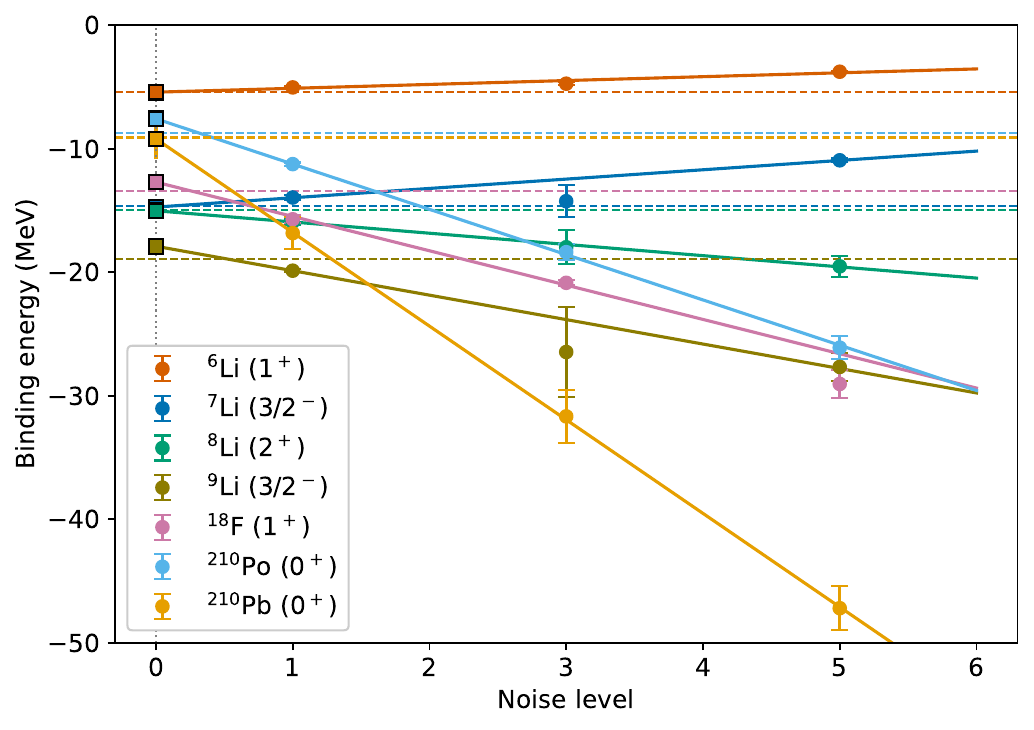}
    \caption{Zero noise extrapolation  performed on the hardware results for all seven nuclei considered in this work with using a linear extrapolation.}
    \label{zne_fig_hardware}
\end{figure*}

\begin{figure*}
    \centering 
    \includegraphics[scale = 0.55]{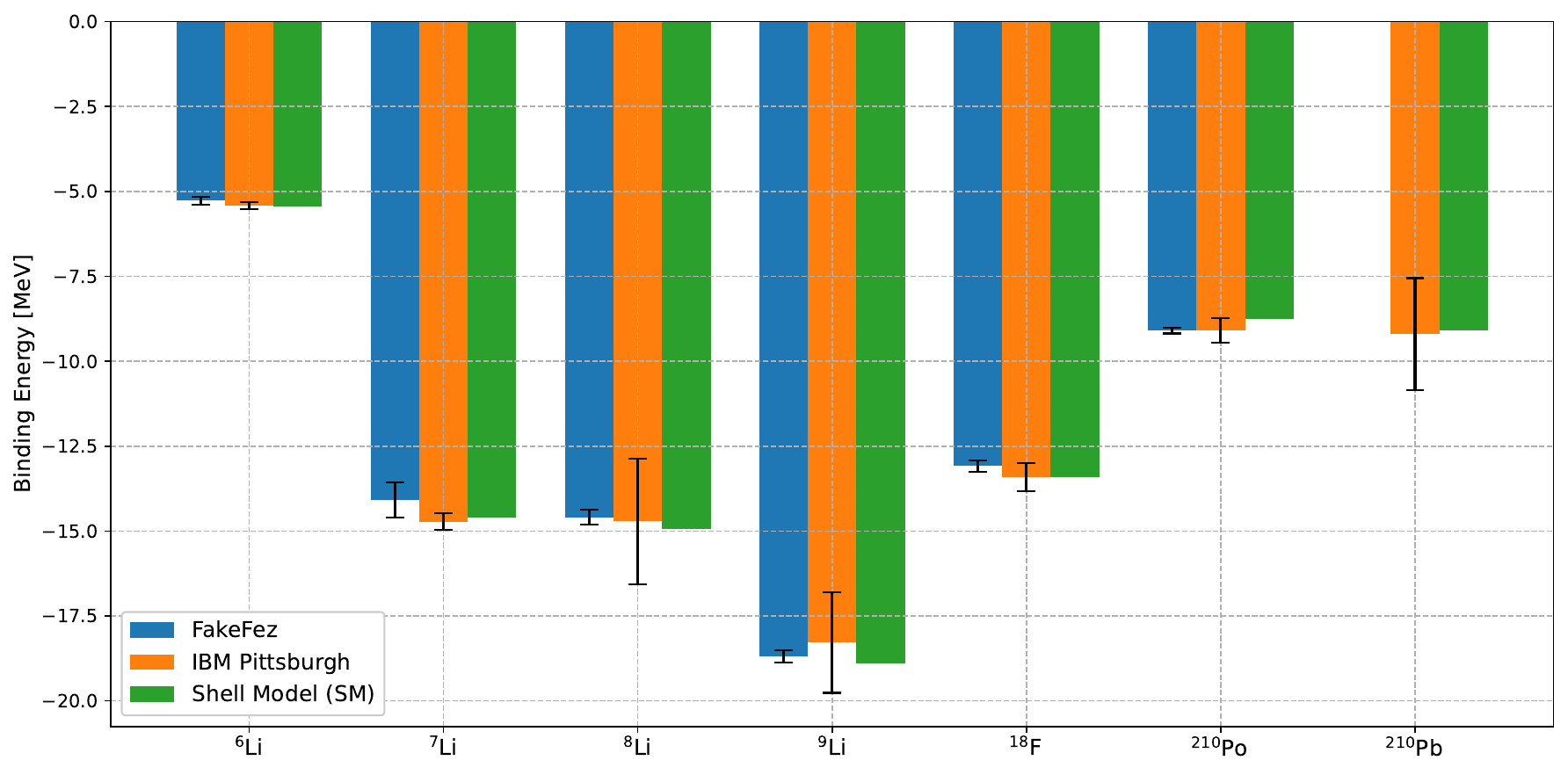}
    \caption{The best error mitigated results from noisy simulator and quantum hardware are compared with the shell model results for the seven nuclei considered. }
    \label{zne_compare}
\end{figure*}

The ZNE is performed for the quantum hardware results for all seven nuclei considered in this work. For each noise factor, all optimized circuits were executed for 10 independent times whose numerical values are represented in third, fourth and fifth columns of \autoref{tab_zne}. \autoref{zne_fig_hardware} shows the linear extrapolation to zero-noise limit using the three sets of hardware results. Like the noisy simulated results, ZNE results corresponding to linear, second degree polynomial, exponential extrapolation and the least percent error are shown in the sixth, seventh, eighth and ninth columns of \autoref{tab_zne}, respectively. The best error mitigated results from nosiy simulation and hardware are compared to the shell model results in \autoref{zne_compare}. Both from \autoref{tab_zne} and \autoref{zne_compare}, it can be seen that the error mitigated binding energies for all seven nuclei are obtained to be within the 4 \% error range compared to the shell model results. The case of $^{210}$Pb is particularly interesting, where the raw hardware results are around 85 \% away from the exact values. However, after applying a linear ZNE, the error mitigated results are only 1.19 \% way from the exact values. 

\begin{table*}
\caption{Numerical results from noisy quantum simulations and hardware executions for each nucleus. For each nucleus, the first row presents data from noisy simulations, while the second row shows results from quantum hardware. Zero-noise extrapolation is performed using one (noise level 3) and two (noise level 5) two-qubit gate folds, fitted with linear, quadratic, and exponential models. For noisy simulations, uncertainties are given as standard deviations from 100 runs for Li circuits and 10 runs for $^{18}$F and $^{210}$Po. Hardware uncertainties are based on 10 independent runs per circuit. Percent errors are calculated relative to the bold-faced reference values, including their associated uncertainties.}
\label{tab_zne}
\resizebox{\textwidth}{!}{%
\begin{tabular}{|c|c|c|c|c|c|c|c|c|}
\hline \\[-8pt]
Isotope & Ref. energy & Noise 1 & Noise 3 & Noise 5 & ZNE (linear)  & ZNE (poly) &ZNE (expo.) & Percent error\\[+2pt]
 \hline \\[-8pt]
$^{6}$Li (1$^+$) 	& -5.437 	& -5.086$\pm$0.083 	& -4.735$\pm$0.093	&  -4.402$\pm$0.079 &  -5.254$\pm$0.100 	& -5.269$\pm$0.197 & \textbf{-5.273$\pm$0.107} & 3.02 \\[+1pt]
& & -5.033 $\pm$ 0.079 & -4.734 $\pm$ 0.105 & -3.764 $\pm$ 0.087 & \textbf{-5.420 $\pm$ 0.098} & -4.933 $\pm$ 0.201 & -5.468 $\pm$ 0.108 & 0.31\\\hline \\[-8pt]
$^{7}$Li (3/2$^-$) & -14.607 & -13.616$\pm$0.221 & -12.741$\pm$0.223 & -11.942$\pm$0.207 & -14.0213$\pm$0.265 & \textbf{-14.082$\pm$0.509} & -14.067$\pm$0.282 & 3.59\\[+1pt]
& & -13.948 $\pm$ 0.196 & -14.247 $\pm$ 1.292 & -10.936 $\pm$ 0.144 & \textbf{-14.724 $\pm$0.247} &  -12.445 $\pm$ 1.650 & -14.844 $\pm$ 0.428 & 0.80\\\hline \\[-8pt]
$^{8}$Li (2$^+$) & -14.926 & \textbf{-14.594$\pm$0.218}& -- & -- & -- & -- & --& 2.22\\[+1pt]
& & -15.926 $\pm$ 0.261 & -17.974 $\pm$ 1.392 & -19.520 $\pm$ 0.861 & -15.022 $\pm$ 0.389 & \textbf{-14.714 $\pm$ 1.848} & -15.133 $\pm$ 0.662 & 1.42\\
\hline \\[-8pt]
$^{9}$Li (3/2$^-$) & -18.906 & \textbf{-18.696$\pm$0.181}& --& -- & -- & -- & -- & 1.46\\
& & -19.872 $\pm$ 0.133 & -26.456 $\pm$ 3.667 & -27.665 $\pm$ 1.137 & -17.894 $\pm$ 0.327 & -14.564 $\pm$ 4.619 & \textbf{-18.277 $\pm$ 1.486} & 3.33\\
\hline \\[-8pt]
$^{18}$F (1$^+$) & -13.413& \textbf{-13.088$\pm$0.178} & -- & -- & -- & -- & -- & 2.42\\[+1pt]
& & -15.710 $\pm$ 0.338 & -20.854 $\pm$ 0.216 & -29.043 $\pm$ 1.108 & -12.704 $\pm$ 0.487 & -14.280 $\pm$ 0.813 & \textbf{-13.418 $\pm$ 0.418} & 0.04\\\hline \\[-8pt]
$^{210}$Po (0$^+$) & -8.762 & -9.438$\pm$0.071 & -10.055$\pm$0.077& -10.770$\pm$0.089 & \textbf{-9.093$\pm$0.089} & -9.166$\pm$0.502 & -9.124$\pm$0.084 & 3.78\\[+1pt]
& & -11.244 $\pm$ 0.129 & -18.373 $\pm$ 0.652 & -26.102 $\pm$ 0.910 & -7.573 $\pm$ 0.253 & -7.904 $\pm$ 0.908 & \textbf{-9.085 $\pm$ 0.359} & 3.69\\\hline \\[-8pt]
$^{210}$Pb (0$^+$) & -9.091 & -- & -- & -- & -- & -- & -- & --\\[+1pt]
& & -16.830 $\pm$ 1.288 & -31.668 $\pm$ 2.125 & -47.197 $\pm$ 1.767 & \textbf{-9.199 $\pm$ 1.647} & -9.670 $\pm$ 3.656 & -13.744 $\pm$ 1.243 & 1.19\\
\hline
\end{tabular}
}
\end{table*}

\section{Summary and Conclusions} 
\label{sect 4}
In this work, we chose to map each possible Slater determinant (SD) of a nucleus within the shell model framework to a qubit, rather than assigning a qubit to each individual single-particle state. Though this way of defining qubits may increase the required number of qubits in some cases, it leads to simpler circuits that are suitable for running on current quantum computers. Firstly, we considered four Li-isotopes $^{6-9}$Li to test single excitation ansatzes defining their ground state. The noisy simulated results from the \textit{FakeFez} backend and the hardware results from $ibm\_pittsburgh$ quantum computer showed that the ground state binding energies are at most 7.5 \% away from the shell model results. Secondly, we made a  comparison between the single excitation Slater determinant basis and double excitation single particle basis ansatzes defining the ground states of $^6$Li and $^{18}$F. The comparison showed a significantly lower resource count for the single excitation ansatz, particularly the two-qubit gates which are one of the major source of error in the NISQ era quantum devices. Moreover, the depth of the single excitation ansatzes is substantially reduced, and the Hamiltonian comprises significantly fewer Pauli terms. Then we considered two heavier nuclei, $^{210}$Po and $^{210}$Pb within the same formalism which were described as 22-qubit and 29-qubit systems, respectively. Finally, we applied the Zero-Noise Extrapolation (ZNE) error mitigation technique using two-qubit gate folding on selected noisy simulation results as well as all hardware-executed results. The best noisy simulated and hardware results
after nosie mitigation are less than 4 \% away from the shell model results for all seven nuclei considered in this wrok.

This method of representing qubits becomes challenging for more complex nuclei, which can quickly exceed current hardware limitations. In such cases, instead of mapping each Slater determinant (SD) to a single qubit, one can encode each SD as a specific multi-qubit state—such as those used in Gray code schemes. However, this approach increases the number of Pauli terms that need to be measured \cite{li_deep_2024}. Conversely, the conventional method of assigning each single-particle state to a qubit leads to a higher number of two-qubit gates. Considering these trade-offs, the way of qubit mapping used in this work remains effective for lighter nuclei and two-nucleon systems across the nuclear chart. As quantum hardware advances toward utility-scale devices with over 100 qubits, strategies that trade increased qubit count for reduced gate complexity and shallower circuit depth offer a promising direction for scalable quantum simulations in nuclear physics.

\section*{ACKNOWLEDGMENTS}
This work is funded and supported by UK STFC grant ST/Y000358/1 and by the UK National Quantum Computer Centre [NQCC200921], which is a UKRI Centre and part of the UK National Quantum Technologies Programme (NQTP).


\bibliography{refs}

\begin{thebibliography}{52}%
\makeatletter
\providecommand \@ifxundefined [1]{%
 \@ifx{#1\undefined}
}%
\providecommand \@ifnum [1]{%
 \ifnum #1\expandafter \@firstoftwo
 \else \expandafter \@secondoftwo
 \fi
}%
\providecommand \@ifx [1]{%
 \ifx #1\expandafter \@firstoftwo
 \else \expandafter \@secondoftwo
 \fi
}%
\providecommand \natexlab [1]{#1}%
\providecommand \enquote  [1]{``#1''}%
\providecommand \bibnamefont  [1]{#1}%
\providecommand \bibfnamefont [1]{#1}%
\providecommand \citenamefont [1]{#1}%
\providecommand \href@noop [0]{\@secondoftwo}%
\providecommand \href [0]{\begingroup \@sanitize@url \@href}%
\providecommand \@href[1]{\@@startlink{#1}\@@href}%
\providecommand \@@href[1]{\endgroup#1\@@endlink}%
\providecommand \@sanitize@url [0]{\catcode `\\12\catcode `\$12\catcode
  `\&12\catcode `\#12\catcode `\^12\catcode `\_12\catcode `\%12\relax}%
\providecommand \@@startlink[1]{}%
\providecommand \@@endlink[0]{}%
\providecommand \url  [0]{\begingroup\@sanitize@url \@url }%
\providecommand \@url [1]{\endgroup\@href {#1}{\urlprefix }}%
\providecommand \urlprefix  [0]{URL }%
\providecommand \Eprint [0]{\href }%
\providecommand \doibase [0]{http://dx.doi.org/}%
\providecommand \selectlanguage [0]{\@gobble}%
\providecommand \bibinfo  [0]{\@secondoftwo}%
\providecommand \bibfield  [0]{\@secondoftwo}%
\providecommand \translation [1]{[#1]}%
\providecommand \BibitemOpen [0]{}%
\providecommand \bibitemStop [0]{}%
\providecommand \bibitemNoStop [0]{.\EOS\space}%
\providecommand \EOS [0]{\spacefactor3000\relax}%
\providecommand \BibitemShut  [1]{\csname bibitem#1\endcsname}%
\let\auto@bib@innerbib\@empty
\bibitem [{\citenamefont {Mayer}\ and\ \citenamefont
  {Jensen}(1955)}]{mayer_jensen}%
  \BibitemOpen
  \bibfield  {author} {\bibinfo {author} {\bibfnamefont {M.~G.}\ \bibnamefont
  {Mayer}}\ and\ \bibinfo {author} {\bibfnamefont {J.~H.~D.}\ \bibnamefont
  {Jensen}},\ }\href@noop {} {\emph {\bibinfo {title} {Elementary {T}heory of
  {N}uclear {S}hell {S}tructure}}}\ (\bibinfo  {publisher} {Wiley},\ \bibinfo
  {address} {New York},\ \bibinfo {year} {1955})\BibitemShut {NoStop}%
\bibitem [{\citenamefont {Gargano}\ \emph {et~al.}(2022)\citenamefont
  {Gargano}, \citenamefont {De~{G}regorio},\ and\ \citenamefont
  {Lenzi}}]{shellmodelmdpi}%
  \BibitemOpen
  \bibinfo {editor} {\bibfnamefont {A.}~\bibnamefont {Gargano}}, \bibinfo
  {editor} {\bibfnamefont {G.}~\bibnamefont {De~{G}regorio}}, \ and\ \bibinfo
  {editor} {\bibfnamefont {S.~M.}\ \bibnamefont {Lenzi}},\ eds.,\ \href
  {\doibase 10.3390/books978-3-0365-9505-4} {\emph {\bibinfo {title} {The
  Nuclear Shell Model 70 Years after Its Advent: Achievements and Prospects}}}\
  (\bibinfo  {publisher} {MDPI},\ \bibinfo {address} {Basel},\ \bibinfo {year}
  {2022})\BibitemShut {NoStop}%
\bibitem [{\citenamefont {Suhonen}(2007)}]{suhonen}%
  \BibitemOpen
  \bibfield  {author} {\bibinfo {author} {\bibfnamefont {J.}~\bibnamefont
  {Suhonen}},\ }\href {\doibase 10.1007/978-3-540-48861-3} {\emph {\bibinfo
  {title} {From {N}ucleons to {N}ucleus}}}\ (\bibinfo  {publisher} {Springer
  Berlin, Heidelberg},\ \bibinfo {year} {2007})\BibitemShut {NoStop}%
\bibitem [{\citenamefont {Dean}\ \emph {et~al.}(2008)\citenamefont {Dean},
  \citenamefont {Hagen}, \citenamefont {Hjorth-Jensen},\ and\ \citenamefont
  {Papenbrock}}]{dean_computational_2008}%
  \BibitemOpen
  \bibfield  {author} {\bibinfo {author} {\bibfnamefont {D.~J.}\ \bibnamefont
  {Dean}}, \bibinfo {author} {\bibfnamefont {G.}~\bibnamefont {Hagen}},
  \bibinfo {author} {\bibfnamefont {M.}~\bibnamefont {Hjorth-Jensen}}, \ and\
  \bibinfo {author} {\bibfnamefont {T.}~\bibnamefont {Papenbrock}},\ }\href
  {\doibase 10.1088/1749-4699/1/1/015008} {\bibfield  {journal} {\bibinfo
  {journal} {Computational Science \& Discovery}\ }\textbf {\bibinfo {volume}
  {1}},\ \bibinfo {pages} {015008} (\bibinfo {year} {2008})}\BibitemShut
  {NoStop}%
\bibitem [{\citenamefont {Andreozzi}\ \emph {et~al.}(2004)\citenamefont
  {Andreozzi}, \citenamefont {Lo~Iudice},\ and\ \citenamefont
  {Porrino}}]{andreozzi_importance_2004}%
  \BibitemOpen
  \bibfield  {author} {\bibinfo {author} {\bibfnamefont {F.}~\bibnamefont
  {Andreozzi}}, \bibinfo {author} {\bibfnamefont {N.}~\bibnamefont
  {Lo~Iudice}}, \ and\ \bibinfo {author} {\bibfnamefont {A.}~\bibnamefont
  {Porrino}},\ }\href {\doibase 10.1134/1.1811187} {\bibfield  {journal}
  {\bibinfo  {journal} {Physics of Atomic Nuclei}\ }\textbf {\bibinfo {volume}
  {67}},\ \bibinfo {pages} {1834} (\bibinfo {year} {2004})}\BibitemShut
  {NoStop}%
\bibitem [{\citenamefont {Shimizu}\ \emph {et~al.}(2019)\citenamefont
  {Shimizu}, \citenamefont {Mizusaki}, \citenamefont {Utsuno},\ and\
  \citenamefont {Tsunoda}}]{shimizu_thick-restart_2019}%
  \BibitemOpen
  \bibfield  {author} {\bibinfo {author} {\bibfnamefont {N.}~\bibnamefont
  {Shimizu}}, \bibinfo {author} {\bibfnamefont {T.}~\bibnamefont {Mizusaki}},
  \bibinfo {author} {\bibfnamefont {Y.}~\bibnamefont {Utsuno}}, \ and\ \bibinfo
  {author} {\bibfnamefont {Y.}~\bibnamefont {Tsunoda}},\ }\href {\doibase
  10.1016/j.cpc.2019.06.011} {\bibfield  {journal} {\bibinfo  {journal}
  {Computer Physics Communications}\ }\textbf {\bibinfo {volume} {244}},\
  \bibinfo {pages} {372} (\bibinfo {year} {2019})}\BibitemShut {NoStop}%
\bibitem [{\citenamefont {Dao}\ and\ \citenamefont
  {Nowacki}(2022)}]{daonowacki}%
  \BibitemOpen
  \bibfield  {author} {\bibinfo {author} {\bibfnamefont {D.~D.}\ \bibnamefont
  {Dao}}\ and\ \bibinfo {author} {\bibfnamefont {F.}~\bibnamefont {Nowacki}},\
  }\href {\doibase 10.1103/PhysRevC.105.054314} {\bibfield  {journal} {\bibinfo
   {journal} {Phys. Rev. C}\ }\textbf {\bibinfo {volume} {105}},\ \bibinfo
  {pages} {054314} (\bibinfo {year} {2022})}\BibitemShut {NoStop}%
\bibitem [{\citenamefont {Stumpf}\ \emph {et~al.}(2016)\citenamefont {Stumpf},
  \citenamefont {Braun},\ and\ \citenamefont
  {Roth}}]{stumpf_importance-truncated_2016}%
  \BibitemOpen
  \bibfield  {author} {\bibinfo {author} {\bibfnamefont {C.}~\bibnamefont
  {Stumpf}}, \bibinfo {author} {\bibfnamefont {J.}~\bibnamefont {Braun}}, \
  and\ \bibinfo {author} {\bibfnamefont {R.}~\bibnamefont {Roth}},\ }\href
  {\doibase 10.1103/PhysRevC.93.021301} {\bibfield  {journal} {\bibinfo
  {journal} {Physical Review C}\ }\textbf {\bibinfo {volume} {93}},\ \bibinfo
  {pages} {021301} (\bibinfo {year} {2016})}\BibitemShut {NoStop}%
\bibitem [{\citenamefont {Launey}\ \emph {et~al.}(2016)\citenamefont {Launey},
  \citenamefont {Dytrych},\ and\ \citenamefont
  {Draayer}}]{launey_symmetry-guided_2016}%
  \BibitemOpen
  \bibfield  {author} {\bibinfo {author} {\bibfnamefont {K.~D.}\ \bibnamefont
  {Launey}}, \bibinfo {author} {\bibfnamefont {T.}~\bibnamefont {Dytrych}}, \
  and\ \bibinfo {author} {\bibfnamefont {J.~P.}\ \bibnamefont {Draayer}},\
  }\href {\doibase 10.1016/j.ppnp.2016.02.001} {\bibfield  {journal} {\bibinfo
  {journal} {Progress in Particle and Nuclear Physics}\ }\textbf {\bibinfo
  {volume} {89}},\ \bibinfo {pages} {101} (\bibinfo {year} {2016})}\BibitemShut
  {NoStop}%
\bibitem [{\citenamefont {Garc\'{\i}a-Ramos}\ \emph {et~al.}(2024)\citenamefont
  {Garc\'{\i}a-Ramos}, \citenamefont {S\'aiz}, \citenamefont {Arias},
  \citenamefont {Lamata},\ and\ \citenamefont {P\'erez-Fern\'andez}}]{QC_rev}%
  \BibitemOpen
  \bibfield  {author} {\bibinfo {author} {\bibfnamefont {J.-E.}\ \bibnamefont
  {Garc\'{\i}a-Ramos}}, \bibinfo {author} {\bibfnamefont {A.}~\bibnamefont
  {S\'aiz}}, \bibinfo {author} {\bibfnamefont {J.~M.}\ \bibnamefont {Arias}},
  \bibinfo {author} {\bibfnamefont {L.}~\bibnamefont {Lamata}}, \ and\ \bibinfo
  {author} {\bibfnamefont {P.}~\bibnamefont {P\'erez-Fern\'andez}},\ }\href
  {\doibase 10.1002/qute.202300219} {\bibfield  {journal} {\bibinfo  {journal}
  {Advanced Quantum Technologies}\ ,\ \bibinfo {pages} {2300219}} (\bibinfo
  {year} {2024})}\BibitemShut {NoStop}%
\bibitem [{\citenamefont {Kiss}\ \emph {et~al.}(2022)\citenamefont {Kiss},
  \citenamefont {Grossi}, \citenamefont {Lougovski}, \citenamefont {Sanchez},
  \citenamefont {Vallecorsa},\ and\ \citenamefont
  {Papenbrock}}]{kiss_quantum_2022}%
  \BibitemOpen
  \bibfield  {author} {\bibinfo {author} {\bibfnamefont {O.}~\bibnamefont
  {Kiss}}, \bibinfo {author} {\bibfnamefont {M.}~\bibnamefont {Grossi}},
  \bibinfo {author} {\bibfnamefont {P.}~\bibnamefont {Lougovski}}, \bibinfo
  {author} {\bibfnamefont {F.}~\bibnamefont {Sanchez}}, \bibinfo {author}
  {\bibfnamefont {S.}~\bibnamefont {Vallecorsa}}, \ and\ \bibinfo {author}
  {\bibfnamefont {T.}~\bibnamefont {Papenbrock}},\ }\href {\doibase
  10.1103/PhysRevC.106.034325} {\bibfield  {journal} {\bibinfo  {journal}
  {Physical Review C}\ }\textbf {\bibinfo {volume} {106}},\ \bibinfo {pages}
  {034325} (\bibinfo {year} {2022})}\BibitemShut {NoStop}%
\bibitem [{\citenamefont {Romero}\ \emph {et~al.}(2022)\citenamefont {Romero},
  \citenamefont {Engel}, \citenamefont {Tang},\ and\ \citenamefont
  {Economou}}]{romero_solving_2022}%
  \BibitemOpen
  \bibfield  {author} {\bibinfo {author} {\bibfnamefont {A.~M.}\ \bibnamefont
  {Romero}}, \bibinfo {author} {\bibfnamefont {J.}~\bibnamefont {Engel}},
  \bibinfo {author} {\bibfnamefont {H.~L.}\ \bibnamefont {Tang}}, \ and\
  \bibinfo {author} {\bibfnamefont {S.~E.}\ \bibnamefont {Economou}},\ }\href
  {\doibase 10.1103/PhysRevC.105.064317} {\bibfield  {journal} {\bibinfo
  {journal} {Physical Review C}\ }\textbf {\bibinfo {volume} {105}},\ \bibinfo
  {pages} {064317} (\bibinfo {year} {2022})}\BibitemShut {NoStop}%
\bibitem [{\citenamefont {Stetcu}\ \emph {et~al.}(2022)\citenamefont {Stetcu},
  \citenamefont {Baroni},\ and\ \citenamefont {Carlson}}]{stetcu}%
  \BibitemOpen
  \bibfield  {author} {\bibinfo {author} {\bibfnamefont {I.}~\bibnamefont
  {Stetcu}}, \bibinfo {author} {\bibfnamefont {A.}~\bibnamefont {Baroni}}, \
  and\ \bibinfo {author} {\bibfnamefont {J.}~\bibnamefont {Carlson}},\ }\href
  {\doibase 10.1103/PhysRevC.105.064308} {\bibfield  {journal} {\bibinfo
  {journal} {Phys. Rev. C}\ }\textbf {\bibinfo {volume} {105}},\ \bibinfo
  {pages} {064308} (\bibinfo {year} {2022})}\BibitemShut {NoStop}%
\bibitem [{\citenamefont {Sarma}\ \emph {et~al.}(2023)\citenamefont {Sarma},
  \citenamefont {Di~Matteo}, \citenamefont {Abhishek},\ and\ \citenamefont
  {Srivastava}}]{sarma_prediction_2023}%
  \BibitemOpen
  \bibfield  {author} {\bibinfo {author} {\bibfnamefont {C.}~\bibnamefont
  {Sarma}}, \bibinfo {author} {\bibfnamefont {O.}~\bibnamefont {Di~Matteo}},
  \bibinfo {author} {\bibfnamefont {A.}~\bibnamefont {Abhishek}}, \ and\
  \bibinfo {author} {\bibfnamefont {P.~C.}\ \bibnamefont {Srivastava}},\ }\href
  {\doibase 10.1103/PhysRevC.108.064305} {\bibfield  {journal} {\bibinfo
  {journal} {Phys. Rev. C}\ }\textbf {\bibinfo {volume} {108}},\ \bibinfo
  {pages} {064305} (\bibinfo {year} {2023})}\BibitemShut {NoStop}%
\bibitem [{\citenamefont {P{\'e}rez-Obiol}\ \emph {et~al.}(2023)\citenamefont
  {P{\'e}rez-Obiol}, \citenamefont {Romero}, \citenamefont {Men{\'e}ndez},
  \citenamefont {Rios}, \citenamefont {Garc{\'i}a-S{\'a}ez},\ and\
  \citenamefont {Juli{\'a}-D{\'i}az}}]{perez-obiol_nuclear_2023}%
  \BibitemOpen
  \bibfield  {author} {\bibinfo {author} {\bibfnamefont {A.}~\bibnamefont
  {P{\'e}rez-Obiol}}, \bibinfo {author} {\bibfnamefont {A.~M.}\ \bibnamefont
  {Romero}}, \bibinfo {author} {\bibfnamefont {J.}~\bibnamefont
  {Men{\'e}ndez}}, \bibinfo {author} {\bibfnamefont {A.}~\bibnamefont {Rios}},
  \bibinfo {author} {\bibfnamefont {A.}~\bibnamefont {Garc{\'i}a-S{\'a}ez}}, \
  and\ \bibinfo {author} {\bibfnamefont {B.}~\bibnamefont
  {Juli{\'a}-D{\'i}az}},\ }\href {\doibase 10.1038/s41598-023-39263-7}
  {\bibfield  {journal} {\bibinfo  {journal} {Scientific Reports}\ }\textbf
  {\bibinfo {volume} {13}},\ \bibinfo {pages} {12291} (\bibinfo {year}
  {2023})}\BibitemShut {NoStop}%
\bibitem [{\citenamefont {Bhoy}\ and\ \citenamefont
  {Stevenson}(2024)}]{bhoy_shell-model_2024}%
  \BibitemOpen
  \bibfield  {author} {\bibinfo {author} {\bibfnamefont {B.}~\bibnamefont
  {Bhoy}}\ and\ \bibinfo {author} {\bibfnamefont {P.}~\bibnamefont
  {Stevenson}},\ }\href {\doibase 10.1088/1367-2630/ad5756} {\bibfield
  {journal} {\bibinfo  {journal} {New J. Phys.}\ }\textbf {\bibinfo {volume}
  {26}},\ \bibinfo {pages} {075001} (\bibinfo {year} {2024})}\BibitemShut
  {NoStop}%
\bibitem [{\citenamefont {Hobday}\ \emph {et~al.}(2025)\citenamefont {Hobday},
  \citenamefont {Stevenson},\ and\ \citenamefont
  {Benstead}}]{hobday_variance_2025}%
  \BibitemOpen
  \bibfield  {author} {\bibinfo {author} {\bibfnamefont {I.}~\bibnamefont
  {Hobday}}, \bibinfo {author} {\bibfnamefont {P.~D.}\ \bibnamefont
  {Stevenson}}, \ and\ \bibinfo {author} {\bibfnamefont {J.}~\bibnamefont
  {Benstead}},\ }\href {\doibase 10.1103/44k6-w3dt} {\bibfield  {journal}
  {\bibinfo  {journal} {Physical Review C}\ }\textbf {\bibinfo {volume}
  {111}},\ \bibinfo {pages} {064321} (\bibinfo {year} {2025})}\BibitemShut
  {NoStop}%
\bibitem [{\citenamefont {Li}\ \emph {et~al.}(2024{\natexlab{a}})\citenamefont
  {Li}, \citenamefont {Baroni}, \citenamefont {Stetcu},\ and\ \citenamefont
  {Humble}}]{li_deep_2024}%
  \BibitemOpen
  \bibfield  {author} {\bibinfo {author} {\bibfnamefont {A.}~\bibnamefont
  {Li}}, \bibinfo {author} {\bibfnamefont {A.}~\bibnamefont {Baroni}}, \bibinfo
  {author} {\bibfnamefont {I.}~\bibnamefont {Stetcu}}, \ and\ \bibinfo {author}
  {\bibfnamefont {T.~S.}\ \bibnamefont {Humble}},\ }\href {\doibase
  10.1140/epja/s10050-024-01286-7} {\bibfield  {journal} {\bibinfo  {journal}
  {The European Physical Journal A}\ }\textbf {\bibinfo {volume} {60}},\
  \bibinfo {pages} {106} (\bibinfo {year} {2024}{\natexlab{a}})}\BibitemShut
  {NoStop}%
\bibitem [{\citenamefont {Carrasco-Codina}\ \emph {et~al.}(2025)\citenamefont
  {Carrasco-Codina}, \citenamefont {Costa}, \citenamefont {Romero},
  \citenamefont {Menéndez},\ and\ \citenamefont
  {Rios}}]{carrascocodina2025comparisonvariationalquantumeigensolvers}%
  \BibitemOpen
  \bibfield  {author} {\bibinfo {author} {\bibfnamefont {M.}~\bibnamefont
  {Carrasco-Codina}}, \bibinfo {author} {\bibfnamefont {E.}~\bibnamefont
  {Costa}}, \bibinfo {author} {\bibfnamefont {A.~M.}\ \bibnamefont {Romero}},
  \bibinfo {author} {\bibfnamefont {J.}~\bibnamefont {Menéndez}}, \ and\
  \bibinfo {author} {\bibfnamefont {A.}~\bibnamefont {Rios}},\ }\href
  {https://arxiv.org/abs/2507.13819} {\enquote {\bibinfo {title} {Comparison of
  variational quantum eigensolvers in light nuclei},}\ } (\bibinfo {year}
  {2025}),\ \Eprint {http://arxiv.org/abs/2507.13819} {arXiv:2507.13819
  [nucl-th]} \BibitemShut {NoStop}%
\bibitem [{\citenamefont {Singh}\ \emph {et~al.}(2025)\citenamefont {Singh},
  \citenamefont {Siwach},\ and\ \citenamefont {Arumugam}}]{nifeeya}%
  \BibitemOpen
  \bibfield  {author} {\bibinfo {author} {\bibfnamefont {N.}~\bibnamefont
  {Singh}}, \bibinfo {author} {\bibfnamefont {P.}~\bibnamefont {Siwach}}, \
  and\ \bibinfo {author} {\bibfnamefont {P.}~\bibnamefont {Arumugam}},\ }\href
  {\doibase 10.1103/bbkf-fjxj} {\bibfield  {journal} {\bibinfo  {journal}
  {Phys. Rev. C}\ }\textbf {\bibinfo {volume} {112}},\ \bibinfo {pages}
  {034320} (\bibinfo {year} {2025})}\BibitemShut {NoStop}%
\bibitem [{\citenamefont {Costa}\ \emph
  {et~al.}(2025{\natexlab{a}})\citenamefont {Costa}, \citenamefont
  {Perez-Obiol}, \citenamefont {Menendez}, \citenamefont {Rios}, \citenamefont
  {Garcia-Saez},\ and\ \citenamefont {Juliá-Díaz}}]{costa_quantum_2025}%
  \BibitemOpen
  \bibfield  {author} {\bibinfo {author} {\bibfnamefont {E.}~\bibnamefont
  {Costa}}, \bibinfo {author} {\bibfnamefont {A.}~\bibnamefont {Perez-Obiol}},
  \bibinfo {author} {\bibfnamefont {J.}~\bibnamefont {Menendez}}, \bibinfo
  {author} {\bibfnamefont {A.}~\bibnamefont {Rios}}, \bibinfo {author}
  {\bibfnamefont {A.}~\bibnamefont {Garcia-Saez}}, \ and\ \bibinfo {author}
  {\bibfnamefont {B.}~\bibnamefont {Juliá-Díaz}},\ }\href {\doibase
  10.21468/SciPostPhys.19.2.062} {\bibfield  {journal} {\bibinfo  {journal}
  {SciPost Physics}\ }\textbf {\bibinfo {volume} {19}},\ \bibinfo {pages} {062}
  (\bibinfo {year} {2025}{\natexlab{a}})}\BibitemShut {NoStop}%
\bibitem [{\citenamefont {Ayral}\ \emph {et~al.}(2023)\citenamefont {Ayral},
  \citenamefont {Besserve}, \citenamefont {Lacroix},\ and\ \citenamefont
  {Ruiz~Guzman}}]{ayral_quantum_2023}%
  \BibitemOpen
  \bibfield  {author} {\bibinfo {author} {\bibfnamefont {T.}~\bibnamefont
  {Ayral}}, \bibinfo {author} {\bibfnamefont {P.}~\bibnamefont {Besserve}},
  \bibinfo {author} {\bibfnamefont {D.}~\bibnamefont {Lacroix}}, \ and\
  \bibinfo {author} {\bibfnamefont {E.~A.}\ \bibnamefont {Ruiz~Guzman}},\
  }\href {\doibase 10.1140/epja/s10050-023-01141-1} {\bibfield  {journal}
  {\bibinfo  {journal} {The European Physical Journal A}\ }\textbf {\bibinfo
  {volume} {59}},\ \bibinfo {pages} {227} (\bibinfo {year} {2023})}\BibitemShut
  {NoStop}%
\bibitem [{\citenamefont {Yuan}\ \emph {et~al.}(2019)\citenamefont {Yuan},
  \citenamefont {Endo}, \citenamefont {Zhao}, \citenamefont {Li},\ and\
  \citenamefont {Benjamin}}]{yuan_theory_2019}%
  \BibitemOpen
  \bibfield  {author} {\bibinfo {author} {\bibfnamefont {X.}~\bibnamefont
  {Yuan}}, \bibinfo {author} {\bibfnamefont {S.}~\bibnamefont {Endo}}, \bibinfo
  {author} {\bibfnamefont {Q.}~\bibnamefont {Zhao}}, \bibinfo {author}
  {\bibfnamefont {Y.}~\bibnamefont {Li}}, \ and\ \bibinfo {author}
  {\bibfnamefont {S.~C.}\ \bibnamefont {Benjamin}},\ }\href {\doibase
  10.22331/q-2019-10-07-191} {\bibfield  {journal} {\bibinfo  {journal}
  {Quantum}\ }\textbf {\bibinfo {volume} {3}},\ \bibinfo {pages} {191}
  (\bibinfo {year} {2019})}\BibitemShut {NoStop}%
\bibitem [{\citenamefont {Cerezo}\ \emph {et~al.}(2022)\citenamefont {Cerezo},
  \citenamefont {Sharma}, \citenamefont {Arrasmith},\ and\ \citenamefont
  {Coles}}]{cerezo_variational_2022}%
  \BibitemOpen
  \bibfield  {author} {\bibinfo {author} {\bibfnamefont {M.}~\bibnamefont
  {Cerezo}}, \bibinfo {author} {\bibfnamefont {K.}~\bibnamefont {Sharma}},
  \bibinfo {author} {\bibfnamefont {A.}~\bibnamefont {Arrasmith}}, \ and\
  \bibinfo {author} {\bibfnamefont {P.~J.}\ \bibnamefont {Coles}},\ }\href
  {\doibase 10.1038/s41534-022-00611-6} {\bibfield  {journal} {\bibinfo
  {journal} {npj Quantum Information}\ }\textbf {\bibinfo {volume} {8}},\
  \bibinfo {pages} {1} (\bibinfo {year} {2022})}\BibitemShut {NoStop}%
\bibitem [{\citenamefont {Fedorov}\ \emph {et~al.}(2022)\citenamefont
  {Fedorov}, \citenamefont {Peng}, \citenamefont {Govind},\ and\ \citenamefont
  {Alexeev}}]{fedorov_vqe_2022}%
  \BibitemOpen
  \bibfield  {author} {\bibinfo {author} {\bibfnamefont {D.~A.}\ \bibnamefont
  {Fedorov}}, \bibinfo {author} {\bibfnamefont {B.}~\bibnamefont {Peng}},
  \bibinfo {author} {\bibfnamefont {N.}~\bibnamefont {Govind}}, \ and\ \bibinfo
  {author} {\bibfnamefont {Y.}~\bibnamefont {Alexeev}},\ }\href {\doibase
  10.1186/s41313-021-00032-6} {\bibfield  {journal} {\bibinfo  {journal}
  {Materials Theory}\ }\textbf {\bibinfo {volume} {6}},\ \bibinfo {pages} {2}
  (\bibinfo {year} {2022})}\BibitemShut {NoStop}%
\bibitem [{\citenamefont {Higgott}\ \emph {et~al.}(2019)\citenamefont
  {Higgott}, \citenamefont {Wang},\ and\ \citenamefont
  {Brierley}}]{higgott_variational_2019}%
  \BibitemOpen
  \bibfield  {author} {\bibinfo {author} {\bibfnamefont {O.}~\bibnamefont
  {Higgott}}, \bibinfo {author} {\bibfnamefont {D.}~\bibnamefont {Wang}}, \
  and\ \bibinfo {author} {\bibfnamefont {S.}~\bibnamefont {Brierley}},\ }\href
  {\doibase 10.22331/q-2019-07-01-156} {\bibfield  {journal} {\bibinfo
  {journal} {Quantum}\ }\textbf {\bibinfo {volume} {3}},\ \bibinfo {pages}
  {156} (\bibinfo {year} {2019})}\BibitemShut {NoStop}%
\bibitem [{\citenamefont {Nakanishi}\ \emph {et~al.}(2019)\citenamefont
  {Nakanishi}, \citenamefont {Mitarai},\ and\ \citenamefont {Fujii}}]{ssvqe}%
  \BibitemOpen
  \bibfield  {author} {\bibinfo {author} {\bibfnamefont {K.~M.}\ \bibnamefont
  {Nakanishi}}, \bibinfo {author} {\bibfnamefont {K.}~\bibnamefont {Mitarai}},
  \ and\ \bibinfo {author} {\bibfnamefont {K.}~\bibnamefont {Fujii}},\ }\href
  {\doibase 10.1103/PhysRevResearch.1.033062} {\bibfield  {journal} {\bibinfo
  {journal} {Phys. Rev. Res.}\ }\textbf {\bibinfo {volume} {1}},\ \bibinfo
  {pages} {033062} (\bibinfo {year} {2019})}\BibitemShut {NoStop}%
\bibitem [{\citenamefont {Li}\ \emph {et~al.}(2024{\natexlab{b}})\citenamefont
  {Li}, \citenamefont {Tao}, \citenamefont {Liang}, \citenamefont {Wu},\ and\
  \citenamefont {Fei}}]{li_full_2024}%
  \BibitemOpen
  \bibfield  {author} {\bibinfo {author} {\bibfnamefont {R.-N.}\ \bibnamefont
  {Li}}, \bibinfo {author} {\bibfnamefont {Y.-H.}\ \bibnamefont {Tao}},
  \bibinfo {author} {\bibfnamefont {J.-M.}\ \bibnamefont {Liang}}, \bibinfo
  {author} {\bibfnamefont {S.-H.}\ \bibnamefont {Wu}}, \ and\ \bibinfo {author}
  {\bibfnamefont {S.-M.}\ \bibnamefont {Fei}},\ }\href {\doibase
  10.1088/1402-4896/ad664c} {\bibfield  {journal} {\bibinfo  {journal} {Physica
  Scripta}\ }\textbf {\bibinfo {volume} {99}},\ \bibinfo {pages} {095207}
  (\bibinfo {year} {2024}{\natexlab{b}})}\BibitemShut {NoStop}%
\bibitem [{\citenamefont {Choi}\ \emph {et~al.}(2021)\citenamefont {Choi},
  \citenamefont {Lee}, \citenamefont {Bonitati}, \citenamefont {Qian},\ and\
  \citenamefont {Watkins}}]{choi_rodeo_2021}%
  \BibitemOpen
  \bibfield  {author} {\bibinfo {author} {\bibfnamefont {K.}~\bibnamefont
  {Choi}}, \bibinfo {author} {\bibfnamefont {D.}~\bibnamefont {Lee}}, \bibinfo
  {author} {\bibfnamefont {J.}~\bibnamefont {Bonitati}}, \bibinfo {author}
  {\bibfnamefont {Z.}~\bibnamefont {Qian}}, \ and\ \bibinfo {author}
  {\bibfnamefont {J.}~\bibnamefont {Watkins}},\ }\href {\doibase
  10.1103/PhysRevLett.127.040505} {\bibfield  {journal} {\bibinfo  {journal}
  {Physical Review Letters}\ }\textbf {\bibinfo {volume} {127}},\ \bibinfo
  {pages} {040505} (\bibinfo {year} {2021})}\BibitemShut {NoStop}%
\bibitem [{\citenamefont {Nigro}\ \emph {et~al.}(2025)\citenamefont {Nigro},
  \citenamefont {Barbieri},\ and\ \citenamefont {Prati}}]{LCU}%
  \BibitemOpen
  \bibfield  {author} {\bibinfo {author} {\bibfnamefont {L.}~\bibnamefont
  {Nigro}}, \bibinfo {author} {\bibfnamefont {C.}~\bibnamefont {Barbieri}}, \
  and\ \bibinfo {author} {\bibfnamefont {E.}~\bibnamefont {Prati}},\ }\href
  {\doibase https://doi.org/10.1002/qute.202400371} {\bibfield  {journal}
  {\bibinfo  {journal} {Advanced Quantum Technologies}\ }\textbf {\bibinfo
  {volume} {8}},\ \bibinfo {pages} {2400371} (\bibinfo {year}
  {2025})}\BibitemShut {NoStop}%
\bibitem [{\citenamefont
  {Robin}(2025)}]{robin2025stabilizeracceleratedquantummanybodygroundstate}%
  \BibitemOpen
  \bibfield  {author} {\bibinfo {author} {\bibfnamefont {C.~E.~P.}\
  \bibnamefont {Robin}},\ }\href {https://arxiv.org/abs/2505.02923} {\enquote
  {\bibinfo {title} {Stabilizer-accelerated quantum many-body ground-state
  estimation},}\ } (\bibinfo {year} {2025}),\ \Eprint
  {http://arxiv.org/abs/2505.02923} {arXiv:2505.02923 [quant-ph]} \BibitemShut
  {NoStop}%
\bibitem [{\citenamefont {Gibbs}\ \emph {et~al.}(2025)\citenamefont {Gibbs},
  \citenamefont {Holmes},\ and\ \citenamefont
  {Stevenson}}]{gibbs_exploiting_2025}%
  \BibitemOpen
  \bibfield  {author} {\bibinfo {author} {\bibfnamefont {J.}~\bibnamefont
  {Gibbs}}, \bibinfo {author} {\bibfnamefont {Z.}~\bibnamefont {Holmes}}, \
  and\ \bibinfo {author} {\bibfnamefont {P.}~\bibnamefont {Stevenson}},\ }\href
  {\doibase 10.1007/s42484-025-00242-y} {\bibfield  {journal} {\bibinfo
  {journal} {Quantum Machine Intelligence}\ }\textbf {\bibinfo {volume} {7}},\
  \bibinfo {pages} {14} (\bibinfo {year} {2025})}\BibitemShut {NoStop}%
\bibitem [{\citenamefont {Miháliková}\ \emph {et~al.}(2025)\citenamefont
  {Miháliková}, \citenamefont {Carlson}, \citenamefont {Neill},\ and\
  \citenamefont {Stetcu}}]{mihalikova_state_2025}%
  \BibitemOpen
  \bibfield  {author} {\bibinfo {author} {\bibfnamefont {I.}~\bibnamefont
  {Miháliková}}, \bibinfo {author} {\bibfnamefont {J.}~\bibnamefont
  {Carlson}}, \bibinfo {author} {\bibfnamefont {D.}~\bibnamefont {Neill}}, \
  and\ \bibinfo {author} {\bibfnamefont {I.}~\bibnamefont {Stetcu}},\ }\href
  {\doibase 10.48550/arXiv.2510.06702} {\enquote {\bibinfo {title} {State
  preparation and symmetries},}\ } (\bibinfo {year} {2025}),\ \bibinfo {note}
  {arXiv:2510.06702}\BibitemShut {NoStop}%
\bibitem [{\citenamefont {Costa}\ \emph
  {et~al.}(2025{\natexlab{b}})\citenamefont {Costa}, \citenamefont
  {Pérez-Obiol}, \citenamefont {Menéndez}, \citenamefont {Rios},
  \citenamefont {García-Sáez},\ and\ \citenamefont
  {Juliá-Díaz}}]{costa2025quasiparticlepairingencodingatomic}%
  \BibitemOpen
  \bibfield  {author} {\bibinfo {author} {\bibfnamefont {E.}~\bibnamefont
  {Costa}}, \bibinfo {author} {\bibfnamefont {A.}~\bibnamefont {Pérez-Obiol}},
  \bibinfo {author} {\bibfnamefont {J.}~\bibnamefont {Menéndez}}, \bibinfo
  {author} {\bibfnamefont {A.}~\bibnamefont {Rios}}, \bibinfo {author}
  {\bibfnamefont {A.}~\bibnamefont {García-Sáez}}, \ and\ \bibinfo {author}
  {\bibfnamefont {B.}~\bibnamefont {Juliá-Díaz}},\ }\href
  {https://arxiv.org/abs/2510.10118} {\enquote {\bibinfo {title} {Quasiparticle
  pairing encoding of atomic nuclei for quantum annealing},}\ } (\bibinfo
  {year} {2025}{\natexlab{b}}),\ \Eprint {http://arxiv.org/abs/2510.10118}
  {arXiv:2510.10118 [nucl-th]} \BibitemShut {NoStop}%
\bibitem [{\citenamefont {Robin}\ and\ \citenamefont
  {Savage}(2023)}]{robin_quantum_2023}%
  \BibitemOpen
  \bibfield  {author} {\bibinfo {author} {\bibfnamefont {C.~E.~P.}\
  \bibnamefont {Robin}}\ and\ \bibinfo {author} {\bibfnamefont {M.~J.}\
  \bibnamefont {Savage}},\ }\href {\doibase 10.1103/PhysRevC.108.024313}
  {\bibfield  {journal} {\bibinfo  {journal} {Physical Review C}\ }\textbf
  {\bibinfo {volume} {108}},\ \bibinfo {pages} {024313} (\bibinfo {year}
  {2023})}\BibitemShut {NoStop}%
\bibitem [{\citenamefont {de~Shalit}\ and\ \citenamefont
  {Talmi}(1963)}]{deshalit_talmi}%
  \BibitemOpen
  \bibfield  {author} {\bibinfo {author} {\bibfnamefont {A.}~\bibnamefont
  {de~Shalit}}\ and\ \bibinfo {author} {\bibfnamefont {I.}~\bibnamefont
  {Talmi}},\ }\href@noop {} {\emph {\bibinfo {title} {Nuclear {S}hell
  {T}heory}}}\ (\bibinfo  {publisher} {Academic Press},\ \bibinfo {address}
  {New York and London},\ \bibinfo {year} {1963})\BibitemShut {NoStop}%
\bibitem [{\citenamefont {Whitehead}(1972)}]{whitehead_numerical_1972}%
  \BibitemOpen
  \bibfield  {author} {\bibinfo {author} {\bibfnamefont {R.~R.}\ \bibnamefont
  {Whitehead}},\ }\href {\doibase 10.1016/0375-9474(72)90278-3} {\bibfield
  {journal} {\bibinfo  {journal} {Nuclear Physics A}\ }\textbf {\bibinfo
  {volume} {182}},\ \bibinfo {pages} {290} (\bibinfo {year}
  {1972})}\BibitemShut {NoStop}%
\bibitem [{\citenamefont {Brown}\ and\ \citenamefont {Rae}(2014)}]{nushellx}%
  \BibitemOpen
  \bibfield  {author} {\bibinfo {author} {\bibfnamefont {B.}~\bibnamefont
  {Brown}}\ and\ \bibinfo {author} {\bibfnamefont {W.}~\bibnamefont {Rae}},\
  }\href {\doibase https://doi.org/10.1016/j.nds.2014.07.022} {\bibfield
  {journal} {\bibinfo  {journal} {Nuclear Data Sheets}\ }\textbf {\bibinfo
  {volume} {120}},\ \bibinfo {pages} {115} (\bibinfo {year}
  {2014})}\BibitemShut {NoStop}%
\bibitem [{\citenamefont {Johnson}\ \emph {et~al.}(2013)\citenamefont
  {Johnson}, \citenamefont {Ormand},\ and\ \citenamefont {Krastev}}]{bigstick}%
  \BibitemOpen
  \bibfield  {author} {\bibinfo {author} {\bibfnamefont {C.~W.}\ \bibnamefont
  {Johnson}}, \bibinfo {author} {\bibfnamefont {W.~E.}\ \bibnamefont {Ormand}},
  \ and\ \bibinfo {author} {\bibfnamefont {P.~G.}\ \bibnamefont {Krastev}},\
  }\href {\doibase https://doi.org/10.1016/j.cpc.2013.07.022} {\bibfield
  {journal} {\bibinfo  {journal} {Computer Physics Communications}\ }\textbf
  {\bibinfo {volume} {184}},\ \bibinfo {pages} {2761} (\bibinfo {year}
  {2013})}\BibitemShut {NoStop}%
\bibitem [{\citenamefont {Jordan}\ and\ \citenamefont {Wigner}(1928)}]{JW}%
  \BibitemOpen
  \bibfield  {author} {\bibinfo {author} {\bibfnamefont {P.}~\bibnamefont
  {Jordan}}\ and\ \bibinfo {author} {\bibfnamefont {E.}~\bibnamefont
  {Wigner}},\ }\href {\doibase 10.1007/BF01331938} {\bibfield  {journal}
  {\bibinfo  {journal} {Zeitschrift für Physik}\ }\textbf {\bibinfo {volume}
  {47}},\ \bibinfo {pages} {631} (\bibinfo {year} {1928})}\BibitemShut
  {NoStop}%
\bibitem [{\citenamefont {Cohen}\ and\ \citenamefont {Kurath}(1965)}]{ckpot}%
  \BibitemOpen
  \bibfield  {author} {\bibinfo {author} {\bibfnamefont {S.}~\bibnamefont
  {Cohen}}\ and\ \bibinfo {author} {\bibfnamefont {D.}~\bibnamefont {Kurath}},\
  }\href {\doibase 10.1016/0029-5582(65)90148-3} {\bibfield  {journal}
  {\bibinfo  {journal} {Nuclear Physics}\ }\textbf {\bibinfo {volume} {73}},\
  \bibinfo {pages} {1} (\bibinfo {year} {1965})}\BibitemShut {NoStop}%
\bibitem [{\citenamefont {Brown}\ and\ \citenamefont {Richter}(2006)}]{usdb}%
  \BibitemOpen
  \bibfield  {author} {\bibinfo {author} {\bibfnamefont {B.~A.}\ \bibnamefont
  {Brown}}\ and\ \bibinfo {author} {\bibfnamefont {W.~A.}\ \bibnamefont
  {Richter}},\ }\href {\doibase 10.1103/PhysRevC.74.034315} {\bibfield
  {journal} {\bibinfo  {journal} {Phys. Rev. C}\ }\textbf {\bibinfo {volume}
  {74}},\ \bibinfo {pages} {034315} (\bibinfo {year} {2006})}\BibitemShut
  {NoStop}%
\bibitem [{\citenamefont {Warburton}\ and\ \citenamefont {Brown}(1991)}]{khpe}%
  \BibitemOpen
  \bibfield  {author} {\bibinfo {author} {\bibfnamefont {E.~K.}\ \bibnamefont
  {Warburton}}\ and\ \bibinfo {author} {\bibfnamefont {B.~A.}\ \bibnamefont
  {Brown}},\ }\href {\doibase 10.1103/PhysRevC.43.602} {\bibfield  {journal}
  {\bibinfo  {journal} {Phys. Rev. C}\ }\textbf {\bibinfo {volume} {43}},\
  \bibinfo {pages} {602} (\bibinfo {year} {1991})}\BibitemShut {NoStop}%
\bibitem [{\citenamefont {Anselme~Martin}\ \emph {et~al.}(2022)\citenamefont
  {Anselme~Martin}, \citenamefont {Simon},\ and\ \citenamefont {Ran\ifmmode
  \check{c}\else \v{c}\fi{}i\ifmmode~\acute{c}\else \'{c}\fi{}}}]{single_ex}%
  \BibitemOpen
  \bibfield  {author} {\bibinfo {author} {\bibfnamefont {B.}~\bibnamefont
  {Anselme~Martin}}, \bibinfo {author} {\bibfnamefont {P.}~\bibnamefont
  {Simon}}, \ and\ \bibinfo {author} {\bibfnamefont {M.~J.}\ \bibnamefont
  {Ran\ifmmode \check{c}\else \v{c}\fi{}i\ifmmode~\acute{c}\else \'{c}\fi{}}},\
  }\href {\doibase 10.1103/PhysRevResearch.4.023190} {\bibfield  {journal}
  {\bibinfo  {journal} {Phys. Rev. Res.}\ }\textbf {\bibinfo {volume} {4}},\
  \bibinfo {pages} {023190} (\bibinfo {year} {2022})}\BibitemShut {NoStop}%
\bibitem [{\citenamefont {Anselmetti}\ \emph {et~al.}(2021)\citenamefont
  {Anselmetti}, \citenamefont {Wierichs}, \citenamefont {Gogolin},\ and\
  \citenamefont {Parrish}}]{double_ex}%
  \BibitemOpen
  \bibfield  {author} {\bibinfo {author} {\bibfnamefont {G.-L.~R.}\
  \bibnamefont {Anselmetti}}, \bibinfo {author} {\bibfnamefont
  {D.}~\bibnamefont {Wierichs}}, \bibinfo {author} {\bibfnamefont
  {C.}~\bibnamefont {Gogolin}}, \ and\ \bibinfo {author} {\bibfnamefont
  {R.~M.}\ \bibnamefont {Parrish}},\ }\href {\doibase 10.1088/1367-2630/ac2cb3}
  {\bibfield  {journal} {\bibinfo  {journal} {New Journal of Physics}\ }\textbf
  {\bibinfo {volume} {23}},\ \bibinfo {pages} {113010} (\bibinfo {year}
  {2021})}\BibitemShut {NoStop}%
\bibitem [{\citenamefont {Powell}(1994)}]{cobyla}%
  \BibitemOpen
  \bibfield  {author} {\bibinfo {author} {\bibfnamefont {M.~J.~D.}\
  \bibnamefont {Powell}},\ }\enquote {\bibinfo {title} {A direct search
  optimization method that models the objective and constraint functions by
  linear interpolation},}\ in\ \href {\doibase 10.1007/978-94-015-8330-5_4}
  {\emph {\bibinfo {booktitle} {Advances in Optimization and Numerical
  Analysis}}}\ (\bibinfo  {publisher} {Springer Netherlands},\ \bibinfo
  {address} {Dordrecht},\ \bibinfo {year} {1994})\ pp.\ \bibinfo {pages}
  {51--67}\BibitemShut {NoStop}%
\bibitem [{\citenamefont {Kraft}(1988)}]{slsqp}%
  \BibitemOpen
  \bibfield  {author} {\bibinfo {author} {\bibfnamefont {D.}~\bibnamefont
  {Kraft}},\ }\href {https://books.google.co.uk/books?id=4rKaGwAACAAJ} {\emph
  {\bibinfo {title} {A Software Package for Sequential Quadratic
  Programming}}},\ Deutsche Forschungs- und Versuchsanstalt f{\"u}r Luft- und
  Raumfahrt K{\"o}ln: Forschungsbericht\ (\bibinfo  {publisher} {Wiss.
  Berichtswesen d. DFVLR},\ \bibinfo {year} {1988})\BibitemShut {NoStop}%
\bibitem [{\citenamefont {Spall}(1992)}]{spsa}%
  \BibitemOpen
  \bibfield  {author} {\bibinfo {author} {\bibfnamefont {J.}~\bibnamefont
  {Spall}},\ }\href {\doibase 10.1109/9.119632} {\bibfield  {journal} {\bibinfo
   {journal} {IEEE Transactions on Automatic Control}\ }\textbf {\bibinfo
  {volume} {37}},\ \bibinfo {pages} {332} (\bibinfo {year} {1992})}\BibitemShut
  {NoStop}%
\bibitem [{\citenamefont {{Qiskit contributors}}(2023)}]{Qiskit}%
  \BibitemOpen
  \bibfield  {author} {\bibinfo {author} {\bibnamefont {{Qiskit
  contributors}}},\ }\href {\doibase 10.5281/zenodo.2573505} {\enquote
  {\bibinfo {title} {Qiskit: An open-source framework for quantum computing},}\
  } (\bibinfo {year} {2023})\BibitemShut {NoStop}%
\bibitem [{\citenamefont {Yoshida}\ \emph {et~al.}(2024)\citenamefont
  {Yoshida}, \citenamefont {Sato}, \citenamefont {Ogata}, \citenamefont
  {Naito},\ and\ \citenamefont {Kimura}}]{kimura}%
  \BibitemOpen
  \bibfield  {author} {\bibinfo {author} {\bibfnamefont {S.}~\bibnamefont
  {Yoshida}}, \bibinfo {author} {\bibfnamefont {T.}~\bibnamefont {Sato}},
  \bibinfo {author} {\bibfnamefont {T.}~\bibnamefont {Ogata}}, \bibinfo
  {author} {\bibfnamefont {T.}~\bibnamefont {Naito}}, \ and\ \bibinfo {author}
  {\bibfnamefont {M.}~\bibnamefont {Kimura}},\ }\href {\doibase
  10.1103/PhysRevC.109.064305} {\bibfield  {journal} {\bibinfo  {journal}
  {Phys. Rev. C}\ }\textbf {\bibinfo {volume} {109}},\ \bibinfo {pages}
  {064305} (\bibinfo {year} {2024})}\BibitemShut {NoStop}%
\bibitem [{\citenamefont {Temme}\ \emph {et~al.}(2017)\citenamefont {Temme},
  \citenamefont {Bravyi},\ and\ \citenamefont {Gambetta}}]{ZNE}%
  \BibitemOpen
  \bibfield  {author} {\bibinfo {author} {\bibfnamefont {K.}~\bibnamefont
  {Temme}}, \bibinfo {author} {\bibfnamefont {S.}~\bibnamefont {Bravyi}}, \
  and\ \bibinfo {author} {\bibfnamefont {J.~M.}\ \bibnamefont {Gambetta}},\
  }\href {\doibase 10.1103/PhysRevLett.119.180509} {\bibfield  {journal}
  {\bibinfo  {journal} {Phys. Rev. Lett.}\ }\textbf {\bibinfo {volume} {119}},\
  \bibinfo {pages} {180509} (\bibinfo {year} {2017})}\BibitemShut {NoStop}%
\bibitem [{\citenamefont {Dumitrescu}\ \emph {et~al.}(2018)\citenamefont
  {Dumitrescu}, \citenamefont {McCaskey}, \citenamefont {Hagen}, \citenamefont
  {Jansen}, \citenamefont {Morris}, \citenamefont {Papenbrock}, \citenamefont
  {Pooser}, \citenamefont {Dean},\ and\ \citenamefont
  {Lougovski}}]{dumitrescu_cloud_2018}%
  \BibitemOpen
  \bibfield  {author} {\bibinfo {author} {\bibfnamefont {E.~F.}\ \bibnamefont
  {Dumitrescu}}, \bibinfo {author} {\bibfnamefont {A.~J.}\ \bibnamefont
  {McCaskey}}, \bibinfo {author} {\bibfnamefont {G.}~\bibnamefont {Hagen}},
  \bibinfo {author} {\bibfnamefont {G.~R.}\ \bibnamefont {Jansen}}, \bibinfo
  {author} {\bibfnamefont {T.~D.}\ \bibnamefont {Morris}}, \bibinfo {author}
  {\bibfnamefont {T.}~\bibnamefont {Papenbrock}}, \bibinfo {author}
  {\bibfnamefont {R.~C.}\ \bibnamefont {Pooser}}, \bibinfo {author}
  {\bibfnamefont {D.~J.}\ \bibnamefont {Dean}}, \ and\ \bibinfo {author}
  {\bibfnamefont {P.}~\bibnamefont {Lougovski}},\ }\href {\doibase
  10.1103/PhysRevLett.120.210501} {\bibfield  {journal} {\bibinfo  {journal}
  {Physical Review Letters}\ }\textbf {\bibinfo {volume} {120}},\ \bibinfo
  {pages} {210501} (\bibinfo {year} {2018})}\BibitemShut {NoStop}%
\end{thebibliography}%

%
%
%
%
%
\end{document}